\documentclass[manyauthors,nocleardouble,COMPASS]{cernphprep}
\usepackage{graphicx}
\usepackage{subfig}
\usepackage[percent]{overpic}
\usepackage[numbers, square, comma, sort&compress]{natbib}

\usepackage{hyperref}

\newcommand{\mrf}[1]{\mbox{$\mathrm{#1}$}}
\newcommand{\mif}[1]{\mbox{$\mathit{#1}$}}

\newcommand{\PicSizeTwoCol}{45ex}

\newcommand{\OPic}[5]{\begin{overpic}[width=#2]{#1}\put(#3,#4){#5}\end{overpic}}

\newcommand{\OPicTwo}[4]{\OPic{#1}{\PicSizeTwoCol}{#2}{#3}{#4}}

\newcommand{\DZ}{D^0}
\newcommand{\DS}{D^*}
\newcommand{\DP}{D^+}
\newcommand{\DM}{D^-}
\newcommand{\barDZ}{\bar{D}^0}
\newcommand{\DSp}{D^{*+}}
\newcommand{\DSm}{D^{*-}}
\newcommand{\DSpm}{D^{* \pm}}
\newcommand{\Kpi}{K\pi}
\newcommand{\Kpipm}{K^+\pi^-}
\newcommand{\Kpimp}{K^-\pi^+}

\newcommand{\ra}{\rightarrow}
\newcommand{\pip}{\pi ^+ }

\newcommand{\ccbar}{c \bar{c}}
\newcommand{\cbar}{ \bar{c}}
\newcommand{\KTS}{K_2 ^* (1430)}
\newcommand{\KThS}{K_3 ^* (1780)}
\newcommand{\NumberDS}{8100\ }
\newcommand{\NumberDZ}{34000\ }  %% FIXME:AZ check the number
\newcommand{\TotalCrossSection}{$1.9 \pm 0.4 $} %% FIXME:AZ check the number
\newcommand{\gs}{\gamma ^* }

\newcommand{\gmass}{\mrf{GeV}}
\newcommand{\mmass}{\mrf{MeV}}
\newcommand{\gom}{\mrf{GeV/\mif{c}}}

\newcommand{\gomt}{\mrf{(GeV/\mif{c})^2}}

\newcommand{\gommt}{\mrf{(GeV/\mif{c})^{-2}}}

\begin{document}

\begin{titlepage}
\PHnumber{2012--339}
\PHdate{November 6, 2012}

\title{$\DS$ and $D$ Meson Production \\ 
in Muon Nucleon Interactions at 160~\gom}

\Collaboration{The COMPASS Collaboration}
\ShortAuthor{The COMPASS Collaboration}

\begin{abstract}
\label{sec:abstract}
The production of $\DS$ and $D$ mesons in inelastic scattering of 160~\gom\
muons off a $^6$LiD target has been investigated with the COMPASS spectrometer
at CERN for $0.003\ \gomt < Q^2 < 10\ \gomt$ and $3~\times~10^{-5}< x_{Bj} <
0.1$. The study is based on \NumberDS events where a $\DZ$ or $\barDZ$ is
detected subsequently to a $\DSp$ or $\DSm$ decay, and on \NumberDZ events,
where only a $\DZ$ or $\barDZ$ is detected. Kinematic distributions of $\DS$,
$D$ and $\KTS$ are given as a function of their energy $E$, transverse momentum
$p_T$, energy fraction $z$, and of the virtual photon variables $\nu$, $Q^2$ and
$x_{Bj}$.  Semi-inclusive differential $\DS$ production cross-sections are
compared with theoretical predictions for $\DS$ production via photon-gluon
fusion into open charm. The total observed production cross-section for $\DSpm $
mesons with laboratory energies between $22$ and $86$ GeV is
(\TotalCrossSection) nb.  Significant cross-section asymmetries are observed
between $\DSp$ and $\DSm$ production for $\nu < 40$ GeV and $z>0.6$.
\end{abstract}

\vfill
\Submitted{(submitted to European Physical Journal C)}
\end{titlepage}

{\pagestyle{empty}
%%%%%%%%%%%%%%%%%%%%%%%%%%%%%%%%%%%%%%%%%%%%%%%%%%%%%%%%%%%%%%%%%%%%%%%%%%%%%%%%%%%%%%%%%%%%%%%%%%%%%%%%%%%%%%%%%%%%%%%
%
% 2012_auththorlist.tex  
%
%%%%%%%%%%%%%%%%%%%%%%%%%%%%%%%%%%%%%%%%%%%%%%%%%%%%%%%%%%%%%%%%%%%%%%%%%%%%%%%%%%%%%%%%%%%%%%%%%%%%%%%%%%%%%%%%%%%%%%%

\section*{The COMPASS Collaboration}
\label{app:collab}
\renewcommand\labelenumi{\textsuperscript{\theenumi}~}
\renewcommand\theenumi{\arabic{enumi}}
\begin{flushleft}
C.~Adolph\Irefn{erlangen},
M.G.~Alekseev\Irefn{triest_i},
V.Yu.~Alexakhin\Irefn{dubna},
Yu.~Alexandrov\Irefn{moscowlpi}\Deceased,
G.D.~Alexeev\Irefn{dubna},
A.~Amoroso\Irefn{turin_u},
A.A.~Antonov\Irefn{dubna},
A.~Austregesilo\Irefnn{cern}{munichtu},
B.~Bade{\l}ek\Irefn{warsaw},
F.~Balestra\Irefn{turin_u}
% ...
\end{flushleft}

%%%%%%%%%%%%%%%%%%%%%%%%%%%%%%%%%%%%%%%%%%%%%%%%%%%%%%%%%%%%%%%%%%%%%%%%%%%%%%%%%%%%%%%%%%%%%%%%%%%%%%%%%%%%%%%%%%%%%%%
%
% institutes
%
%%%%%%%%%%%%%%%%%%%%%%%%%%%%%%%%%%%%%%%%%%%%%%%%%%%%%%%%%%%%%%%%%%%%%%%%%%%%%%%%%%%%%%%%%%%%%%%%%%%%%%%%%%%%%%%%%%%%%%%

\begin{Authlist}
\item \Idef{bielefeld}{Universit\"at Bielefeld, Fakult\"at f\"ur Physik, 33501 Bielefeld, Germany\Arefs{f}}
\item \Idef{bochum}{Universit\"at Bochum, Institut f\"ur Experimentalphysik, 44780 Bochum, Germany\Arefs{f}}
\item \Idef{bonniskp}{Universit\"at Bonn, Helmholtz-Institut f\"ur  Strahlen- und Kernphysik, 53115 Bonn, Germany\Arefs{f}}
\item \Idef{bonnpi}{Universit\"at Bonn, Physikalisches Institut, 53115 Bonn, Germany\Arefs{f}}
\item \Idef{brno}{Institute of Scientific Instruments, AS CR, 61264 Brno, Czech Republic\Arefs{g}}
\item \Idef{calcutta}{Matrivani Institute of Experimental Research \& Education, Calcutta-700 030, India\Arefs{h}}
\item \Idef{dubna}{Joint Institute for Nuclear Research, 141980 Dubna, Moscow region, Russia\Arefs{i}}
\item \Idef{erlangen}{Universit\"at Erlangen--N\"urnberg, Physikalisches Institut, 91054 Erlangen, Germany\Arefs{f}}
\item \Idef{freiburg}{Universit\"at Freiburg, Physikalisches Institut, 79104 Freiburg, Germany\Arefs{f}}
\item \Idef{cern}{CERN, 1211 Geneva 23, Switzerland}
\item \Idef{liberec}{Technical University in Liberec, 46117 Liberec, Czech Republic\Arefs{g}}
\item \Idef{lisbon}{LIP, 1000-149 Lisbon, Portugal\Arefs{j}}
\item \Idef{mainz}{Universit\"at Mainz, Institut f\"ur Kernphysik, 55099 Mainz, Germany\Arefs{f}}
\item \Idef{miyazaki}{University of Miyazaki, Miyazaki 889-2192, Japan\Arefs{k}}
\item \Idef{moscowlpi}{Lebedev Physical Institute, 119991 Moscow, Russia}
\item \Idef{munichlmu}{Ludwig-Maximilians-Universit\"at M\"unchen, Department f\"ur Physik, 80799 Munich, Germany\Arefs{f}\Arefs{l}}
\item \Idef{munichtu}{Technische Universit\"at M\"unchen, Physik Department, 85748 Garching, Germany\Arefs{f}\Arefs{l}}
\item \Idef{nagoya}{Nagoya University, 464 Nagoya, Japan\Arefs{k}}
\item \Idef{praguecu}{Charles University in Prague, Faculty of Mathematics and Physics, 18000 Prague, Czech Republic\Arefs{g}}
\item \Idef{praguectu}{Czech Technical University in Prague, 16636 Prague, Czech Republic\Arefs{g}}
\item \Idef{protvino}{State Research Center of the Russian Federation, Institute for High Energy Physics, 142281 Protvino, Russia}
\item \Idef{saclay}{CEA IRFU/SPhN Saclay, 91191 Gif-sur-Yvette, France}
\item \Idef{telaviv}{Tel Aviv University, School of Physics and Astronomy, 69978 Tel Aviv, Israel\Arefs{m}}
\item \Idef{triest_i}{Trieste Section of INFN, 34127 Trieste, Italy}
\item \Idef{triest}{University of Trieste, Department of Physics and Trieste Section of INFN, 34127 Trieste, Italy}
\item \Idef{triestictp}{Abdus Salam ICTP and Trieste Section of INFN, 34127 Trieste, Italy}
\item \Idef{turin_u}{University of Turin, Department of Physics and Torino Section of INFN, 10125 Turin, Italy}
\item \Idef{turin_i}{Torino Section of INFN, 10125 Turin, Italy}
\item \Idef{turin_p}{University of Eastern Piedmont, 15100 Alessandria,  and Torino Section of INFN, 10125 Turin, Italy}
\item \Idef{warsaw}{National Centre for Nuclear Research and University of Warsaw, 00-681 Warsaw, Poland\Arefs{n} }
\item \Idef{warsawtu}{Warsaw University of Technology, Institute of Radioelectronics, 00-665 Warsaw, Poland\Arefs{n} }
\item \Idef{yamagata}{Yamagata University, Yamagata, 992-8510 Japan\Arefs{k} }
\end{Authlist}
%%%%%%%%%%%%%%%%%%%%%%%%%%%%%%%%%%%%%%%%%%%%%%%%%%%%%%%%%%%%%%%%%%%%%%%%%%%%%%%%%%%%%%%%%%%%%%%%%%%%%%%%%%%%%%%%%%%%%%%
%
% Notes
%
%%%%%%%%%%%%%%%%%%%%%%%%%%%%%%%%%%%%%%%%%%%%%%%%%%%%%%%%%%%%%%%%%%%%%%%%%%%%%%%%%%%%%%%%%%%%%%%%%%%%%%%%%%%%%%%%%%%%%%%
\vspace*{-\baselineskip}\renewcommand\theenumi{\alph{enumi}}
\begin{Authlist}
\item \Adef{a}{Also at IST, Universidade T\'ecnica de Lisboa, Lisbon, Portugal}
\item \Adef{bb}{Supported by the DFG Research Training Group Programme 1102  ``Physics at Hadron Accelerators''}
\item \Adef{b}{Also at Chubu University, Kasugai, Aichi, 487-8501 Japan\Arefs{k}}
\item \Adef{c}{Also at KEK, 1-1 Oho, Tsukuba, Ibaraki, 305-0801 Japan}
\item \Adef{d}{On leave of absence from JINR Dubna}
\item \Adef{y}{present address: National Science Foundation, 4201 Wilson Boulevard, Arlington, VA 22230, United States}
\item \Adef{x}{present address: RWTH Aachen University, III. Physikalisches Institut, 52056 Aachen, Germany}
\item \Adef{e}{Also at GSI mbH, Planckstr.\ 1, D-64291 Darmstadt, Germany}
\item \Adef{f}{Supported by the German Bundesministerium f\"ur Bildung und Forschung}
\item \Adef{g}{Supported by Czech Republic MEYS Grants ME492 and LA242}
\item \Adef{h}{Supported by SAIL (CSR), Govt.\ of India}
\item \Adef{i}{Supported by CERN-RFBR Grants 08-02-91009}
\item \Adef{j}{\raggedright Supported by the Portuguese FCT - Funda\c{c}\~{a}o para a Ci\^{e}ncia e Tecnologia, COMPETE and QREN, Grants CERN/FP/109323/2009, CERN/FP/116376/2010 and CERN/FP/123600/2011}
\item \Adef{k}{Supported by the MEXT and the JSPS under the Grants No.18002006, No.20540299 and No.18540281; Daiko Foundation and Yamada Foundation}
\item \Adef{l}{Supported by the DFG cluster of excellence `Origin and Structure of the Universe' (www.universe-cluster.de)}
\item \Adef{m}{Supported by the Israel Science Foundation, founded by the Israel Academy of Sciences and Humanities}
\item \Adef{n}{Supported by the Polish NCN Grant DEC-2011/01/M/ST2/02350}
\item [{\makebox[2mm][l]{\textsuperscript{*}}}] Deceased
\end{Authlist}

\clearpage
}

\maketitle

\section{Introduction}
\label{sec:introduction}

The production of $D$ mesons in inelastic scattering of 160~\gom\ muons on
nucleons $ \mu N \ra \mu ^{\prime} D X $ is assumed to be dominated by a
process where the exchanged virtual photon $\gs$ fuses with a gluon $g$ into a
charm anti-charm quark pair, $\gs g \ra \ccbar $. The cross-section $ \sigma
^{\gs g \ra \ccbar }$ of this photon-gluon fusion (PGF) process and its
dependence on the relative polarization of photon and gluon can be calculated in
perturbative QCD \cite{PGFTH11,PGFTH12,PGFTH1,PGFTH13,PGFTH2,PGFTH3,PGFTH4,
PGFTH5}.  Thus, using polarized muons and polarized nucleons, a measurement of
the photon nucleon cross-section asymmetry $\Delta \sigma ^{\gs N \ra \ccbar X}
/ \sigma ^{\gs N \ra \ccbar X}$ allows the determination of the gluon
polarization $\Delta g/g$ in the nucleon. With this objective, open charm
production has been studied in the COMPASS experiment at CERN for longitudinally
polarized muons interacting with longitudinally polarized deuterons. The
incoming muon energy of 160 GeV was chosen, since the cross-section difference
$\Delta \sigma ^{\gs N \ra \ccbar X} $ for parallel and anti-parallel spins of
photon and nucleon reaches a maximum for virtual photon energies around 80 GeV
according to most models for the gluon helicity distribution function $\Delta g
(x_{Bj},Q^2)$, and the polarization transfer from muon to virtual photon is
large in the relevant photon energy range.

Final states, where the decays $\DZ \ra \Kpimp$ or $\DSp \ra \DZ \pip \ra \Kpimp
\pip $ or the charge conjugate decays are detected, are chosen in order to
achieve the best possible combination of mass resolution, signal-over-background
ratio and signal statistics. Based on samples of events with these final states,
extracted from data taken during the years 2002-2006, COMPASS has published
results for $\langle \Delta g/g \rangle$ \cite{LETTER}.

The photon-gluon cross-section asymmetry $a_{LL}$ = $\Delta \sigma ^{\gs g \ra
\ccbar} / \sigma ^{\gs g \ra \ccbar}$ needed for extracting $\langle \Delta g/g
\rangle$ is estimated making two assumptions: only PGF contributes as calculated
in leading order QCD and charm and anti-charm quarks hadronize independently of
the target polarization. The parton kinematics are estimated event-by-event on
the basis of the observed 3-momentum of the $\DZ$ meson and the momentum
difference of the incoming and the scattered muon using a parametrisation
based on the Monte Carlo event generator AROMA with default charm quark
fragmentation \cite{AROMA}.
 
However, production mechanisms other than PGF with standard charm quark
fragmentation may contribute to the observed events with charmed mesons. The
interaction of the virtual photon with intrinsic charm \cite{INTRINSIC2,
INTRINSIC3, INTRINSIC4} is one possible competing mechanism. The associated
production of $\Lambda _c \barDZ$ \cite{ASSOCIATE} or, more generally,
asymmetric hadronization of $c$ and $\cbar$ like in the Dual Parton model with a
meson and a baryon string \cite{DUAL, NA14} may play an important role in some
regions of phase space. A study of the phase space distributions and
semi-inclusive differential cross-sections of the $\DZ$ and $\DS$ mesons within
the acceptance of the COMPASS spectrometer may yield information about the
contributions of different production mechanisms \cite{Zvyagin}.

At HERA, i.e.\ at much larger center-of-mass energy, charm electro-production has
been studied in detail by the H1 and ZEUS Collaborations, see \cite{H1,ZEUS,LIP}
and references therein. In addition to the PGF other production mechanisms also
contribute in this case, like gluon-gluon fusion to $\ccbar$ from a resolved
photon. The hadronizations of $c$ and $\cbar$ can more safely be assumed to be
independent. COMPASS covers a complementary kinematic region with virtual photon
energies in the range from threshold at about 30 GeV up to 140 GeV in the
laboratory frame. Prior to COMPASS, this energy range was covered only by the
EMC experiment \cite{EMC}, which collected about 90 $\DZ$ meson events produced
by deep inelastic scattering of 240 or 280 GeV muons on hydrogen and deuterium
targets for a study of the charm production mechanism. Only one charm
photo-production experiment explored the region close to threshold \cite{E687},
while two concentrated on high energy photons \cite{TPCOLL, NA14}.
 
The present article shows details of the phase space distributions of $D$~meson
production as a function of various kinematic variables: the energy $\nu =
E_{\mu} - E_{\mu}^{\prime}$ of the exchanged virtual photon $\gamma ^*$ 
(assuming single photon exchange) with four momentum $q=p-p^{\prime}$, the
inelasticity $y = \nu / E_{\mu}$ of the 
event, the negative invariant $\gamma ^*$ mass squared $Q^2 = -q^2=-(p-
p^{\prime})^2$ and the Bjorken scaling variable $x_{Bj} = Q^2/2P \cdot q= Q^2/
2M_p \nu $. Here
$E_{\mu} $ and $ E _{\mu}^{\prime}$ are the laboratory energies, $p$ and
$p^{\prime}$ the 4-momentum vectors of the incoming and scattered muon
respectively, $P$ is the 4-momentum of the target nucleon and $M_p$ is the proton mass.

In order to describe both $\DS$ and $\DZ$ meson production, the following
kinematic variables are used: the transverse momentum $p_T$ of the $K\pi$ pair
(from the $\DZ$ decay) with respect to the $\gamma ^*$ direction, the $\DZ$ energy $E$
in the laboratory system and the energy fraction $z = E / \nu$.
 
The outline of the paper is as follows: after a brief overview of the
experimental set-up (section \ref{sec:setup}) and a detailed description of the
data selection procedure (section \ref{sec:data selection}), the methods of
signal extraction are described in section \ref{sec:signal}. The kinematic
distributions of events from signal and background regions are shown in
section~\ref{kinedis}. They are based on the entire available data sample
collected during the years 2002-2006, and are not corrected for acceptance. The
purpose of this section is to compare the distributions of open charm to those
of background events. Section \ref{sec:acceptance} describes the acceptance
correction and luminosity calculation needed to extract the total and
differential semi-inclusive cross sections for charm production. They are
performed for the 2004 data only. In section \ref{sec:cross section}, the
differential cross-sections as a function of the various kinematic variables and
the total cross-section obtained by integration are shown and compared with
available theoretical (AROMA) predictions for the production of $D$ mesons by
PGF. Significant differences between $\DSp $ and $\DSm$ meson production are
observed for the acceptance-corrected data from 2004.  A statistically more
precise comparison of $\DSp $ and $\DSm$ production is based on the entire data
sample (2002-2006).  Particle-antiparticle asymmetries are determined under the
assumption, which is verified for the 2004 data, that the $\DSp $ and $\DSm$
acceptances are equal to a good approximation.

\section{Experimental setup}
\label{sec:setup}

The data were taken using the COMPASS spectrometer situated at the M2
beam line at the CERN Super Proton Synchrotron. A detailed
description of the COMPASS spectrometer can be found in Ref.~\cite{DETECTOR}.

The momentum of the positive muon beam is about 160~\gom\ with a
spread of 5$\%$. The momentum of each incoming muon is measured with a
precision of $\Delta p / p < 1 \% $ in the beam momentum spectrometer
located upstream of the experimental hall, and its direction is
measured with a precision of 30 $\mu\mrf{rad}$ with a detector telescope in
front of the target.

The (polarized) $^6$LiD target consisted of two 60 cm long cells during the
years 2002-2004 and of 3 cells with a total length of 120 cm in 2006. The
polarization is reversed regularly such that the products of integrated
luminosities times acceptance are equal for both polarizations. The sum of both
corresponds to the unpolarized case. Hence unpolarized distributions, 
which are the subject of the present analysis, are obtained
from the sum of all data. The target is housed in a superconducting solenoid
magnet, which determines the angular acceptance of the spectrometer.  The acceptance
in the polar angle, measured at the upstream edge of the target, was 70 mrad
in 2002-2004, while with the new target magnet in 2006 it was increased
to 180 mrad.

The 2-stage spectrometer is designed to reconstruct and identify the scattered
muon and produced hadrons over a wide momentum range. It contains a large angle
(LAS) and a small angle (SAS) part, each part equipped with a dipole magnet.
Tracking detectors are located in front and behind each magnet, and
electromagnetic and hadron calorimeters behind. The LAS covers polar laboratory
angles from about 15 mrad up to 70 mrad in 2002-2004 and, with the new target
magnet, up to 180 mrad in 2006. The SAS covers the polar laboratory angles below 20
mrad.

Particle tracking is done with a large variety of tracking detectors: several
stations of silicon microstrip detectors, scintillating fiber detectors, high
resolution micromesh gaseous chambers, gas electron-multi\-plier chambers, drift
chambers, large area straw drift chambers, multiwire proportional chambers and
muon drift tubes. The scattered muons are identified downstream of additional
hadron absorbers placed behind the hadron calorimeters. Charged hadrons are identified
by a Ring Imaging Cherenkov detector (RICH) in the LAS.

The trigger system \cite{TRIGGER} uses hodoscope and calorimeter information
to select inelastic muon interactions with minimum bias. The overall trigger
and muon track reconstruction efficiency is in the range 60\% to 80\% for most
of the kinematic region covered by COMPASS.

\section{Data selection}
\label{sec:data selection}

The total number of events with an incoming muon ($140\ \gom < p_\mu < 180\
\gom$) and a scattered muon from a common vertex is $5.2 \times 10^9$, which
corresponds to an integrated luminosity of about 2.8 \mrf{fb^{-1}}. This sample
is used to search for $\DZ (\barDZ)$ and $\DSpm$ mesons. A fiducial volume
cut makes sure that the extrapolated incoming muon trajectory traverses all
target cells and that the primary vertex is located within the volume of one of
the target cells.

Since the COMPASS experiment uses a large solid target, the detection of a
secondary decay vertex, which is a standard method in open charm detection, is
excluded and the selection of $\DZ (\barDZ)$ and $\DSpm$ mesons relies on
requirements on event kinematics and particle identification. The event
selection is the same as used in the previous COMPASS open charm publication
\cite{LETTER} except for stricter requirements on the selection
of the incoming muon.

Cuts used to select $\DZ$ originating from the decay of a $\DS$ (referred to as
$\DS$ or `tagged' $\DZ$ sample) slightly differ from those used to select
directly produced $\DZ$ mesons (referred to as `untagged' $\DZ$ sample). An
event from the untagged $\DZ$ sample contains at least one candidate for the
2-body decay $\DZ \rightarrow K^-\pi^+$ or its charge conjugate (c.c.), while in
the tagged $\DZ$ sample a slow pion from the decay chain $\DSp \rightarrow \DZ
\pi^+_s \rightarrow K^-\pi^+\pi^+_s$ (or c.c) has to be present in addition.

Particles are identified by using the RICH. All tracks with momentum measured in
one or both spectrometer stages and falling within the geometrical acceptance of
the RICH are used to calculate the likelihoods (LKs) that the Cherenkov photons
detected by the RICH are due to electron, muon, pion, kaon, proton, or
background. The LK for a specific particle is calculated only if the particle
velocity is above the threshold for the emission of Cherenkov photons in the
radiator gas. This threshold depends on the refractive index that is extracted
from the data on a run-by-run basis. For pions, kaons, and protons, this gives an
average momentum threshold of 2.5, 8.9 and 16.9~\gom\ respectively. At large
momenta pions and kaons cannot be efficiently separated, thus it is required
that the momentum of the particle is below 50~\gom. In the tagged $\DZ$ sample,
due to the small mass difference between the $\DS (2010)$ and the $\DZ (1865)$,
only a limited energy is available for the pion produced in the $\DS \rightarrow
\DZ \pi_s$ decay. In this case, the $\pi_s$ candidate must not have been
identified as an electron by the RICH. Details on the LK requirements and
the use of the RICH information can be found in Ref.~\cite{Zvyagin}.

For untagged $\DZ$, the following cuts are applied to the $K^-\pi^+$ and
$K^+\pi^-$ combinations:  $p_\pi > 7$ \gom, $z >
0.2$, $| M (K \pi) - M (\DZ) | < 700$ and $| \cos \theta_K | < 0.65$, where
$\theta_K$ is the decay kaon angle in 
the $\DZ$ center-of-mass system with respect to the $\DZ$ direction of flight.

\begin{figure}
\centering
\includegraphics[width=55ex]{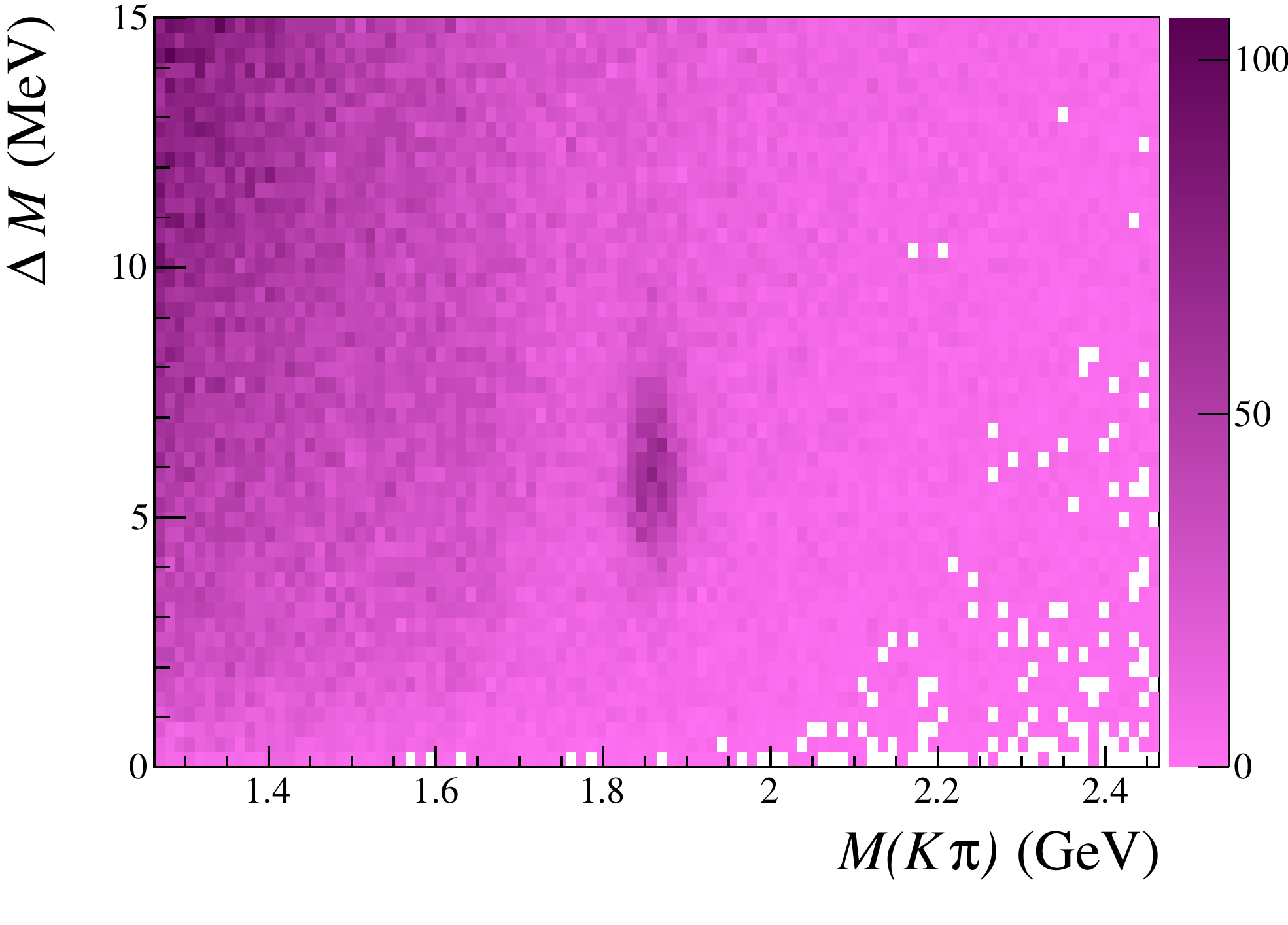}
\caption{
Scatter plot for $\DS$ candidates before applying the $\Delta M$ cut. 
Vertical axis: $\Delta M$; horizontal axis: $M(K\pi)$. 
The accumulation of events around the $\DZ$ nominal mass of 1.864~\gmass\
and  $\Delta M=$ 6.1 \mmass\ corresponds to the decay sequence 
$\DS \rightarrow \pi_s \DZ \rightarrow \pi_s (K\pi)$. \label{fig:DsTag}}
\end{figure}

For the tagged $\DZ$, the $K \pi \pi_s$ invariant mass is calculated only if the
$K\pi$ system has an invariant mass in the range $| M (K \pi) - M (\DZ) | < 700$
\mmass. The distribution of $\Delta M = M(K\pi\pi_s) - M(K\pi) - M(\pi)$ as a
function of $M(K \pi)$ is shown in Fig.~\ref{fig:DsTag}. Here a clear spot for
the $\DS$ is visible at $\Delta M \sim 6$ \mmass\ in the region of the $\DZ$
mass. The cut $3.2\ \mmass < \Delta M < 8.9\ \mmass$ improves
the $\DZ$ signal with respect to the combinatorial background by more than an
order of magnitude. The $K \pi$ system is also required to have $z > 0.2$
and $| \cos \theta_K | < 0.9$. 

These sets of cuts define the untagged and tagged $\DZ$ samples, i.e.\ the $\DZ$
and $\DS$ candidates.

\begin{figure}
\centering
\includegraphics[width=55ex]{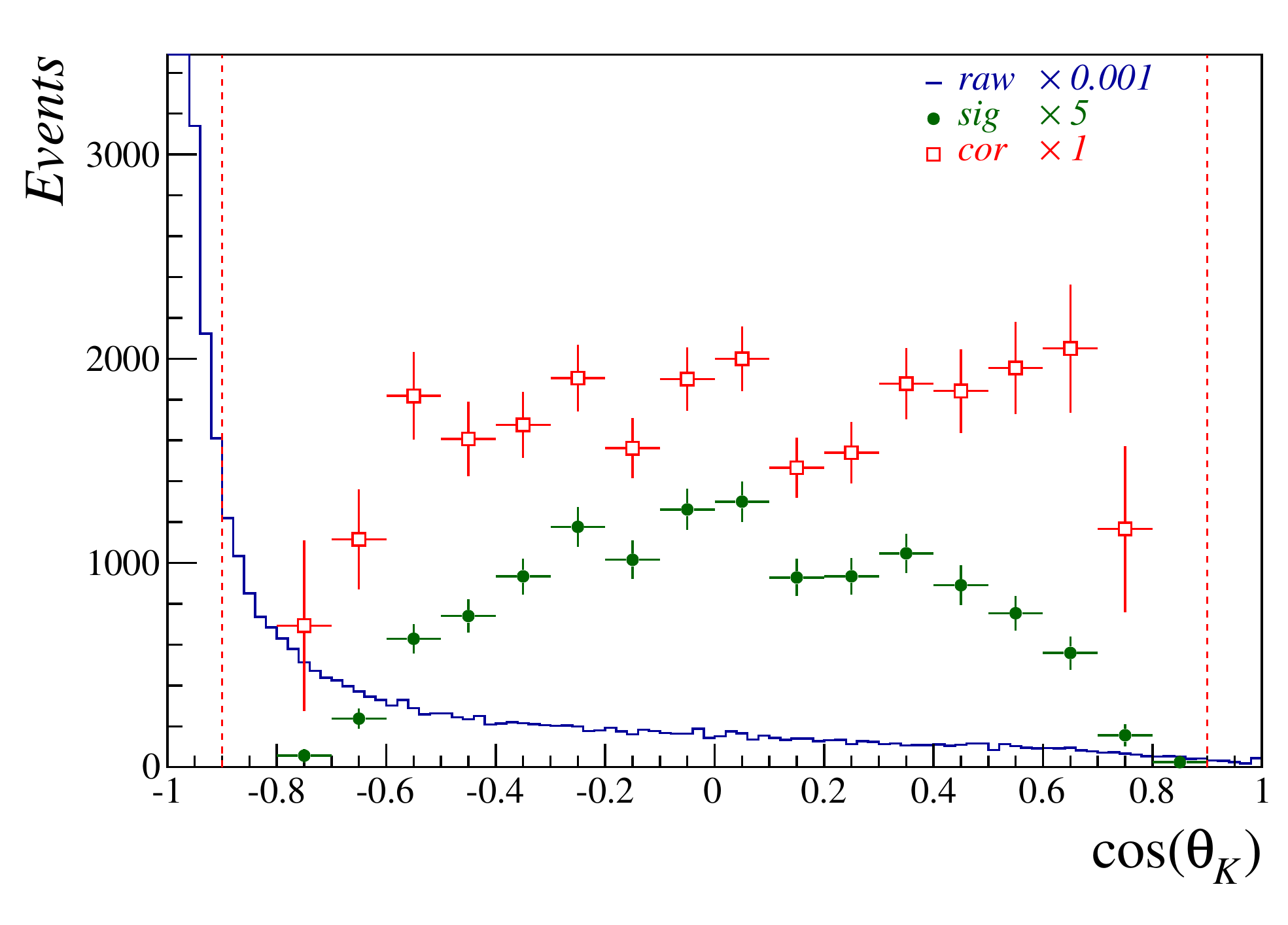}
\caption{Distribution of $\cos\theta_K$ in the $K \pi$ rest frame 
for (mostly) background combinations (scaled by 0.001, solid line), for the $\DZ$
signal region (scaled by 5, full circles) and for the acceptance corrected $\DZ$
signal (open squares). The dashed lines correspond to the $| \cos \theta_K | <
0.9$ cut. The $\DZ$ signal is from the 2004 tagged $\DZ$ sample. \label{fig:cut_raw_sig}}  
\end{figure}

The $\cos \theta_K$ distribution is the only distribution where a safe
theoretical prediction can be made.  The uncorrected $\cos \theta_K$
distribution of $K \pi$ events before any mass cuts, i.e.\ mostly background,
shown in Fig.~\ref{fig:cut_raw_sig} is strongly peaked towards $\cos \theta_K=-1$.
For signal events, the $\cos \theta_K$ distribution should be flat
after acceptance correction since the $\DZ$ has spin 0.  This expectation for
the $\DZ$ is confirmed in Fig.~\ref{fig:cut_raw_sig} where, for the tagged $\DZ$
sample, the distribution for $\DZ$ is shown before and after acceptance
correction (the method of signal extraction and the correction for the
acceptance will be described in secs.~\ref{sec:signal} and~\ref{sec:acceptance}).

The so-called ambiguity cut applied in Ref.~\cite{LETTER} is not applied
in the present analysis. This cut discards an event if two $\DZ$ or
$\barDZ$ meson candidates are found within the mass window of $\pm $700 \mmass\ and
removes a significant number of good events. However, the
probability to find two $\DZ$ (or two $\barDZ$) mesons in the signal peak is
practically zero. Hence the present analysis, which extracts the number of
signal events from fitting separately the $\DZ$ and $\barDZ$ peaks, does not
suffer from this ambiguity. 

In the mass window of $\pm 700$ \mmass\ around the nominal $\DZ$ mass, the tagged
$\DZ$ sample consists of $160\times 10^3$ neutral $\Kpi$ combinations. In order
to avoid overlapping samples, at this stage, the $\DS$ candidates are taken out
of the untagged $\DZ$ sample. In the same mass window, the untagged $\DZ$ sample
comprises $17 \times 10^6$ neutral $\Kpi$ combinations.

\begin{figure}
\centering
\OPicTwo{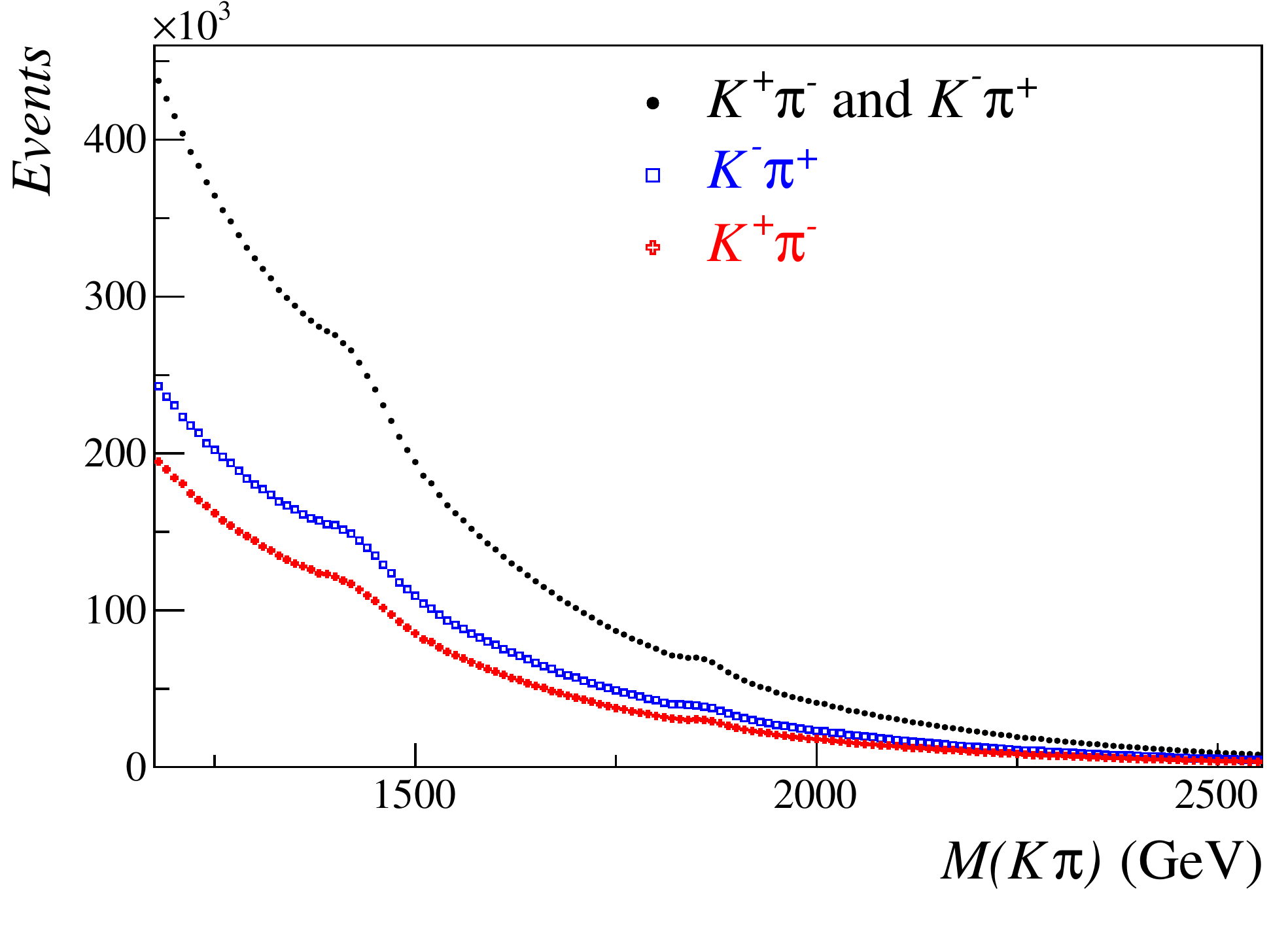}{75}{40}{(a)}
\OPicTwo{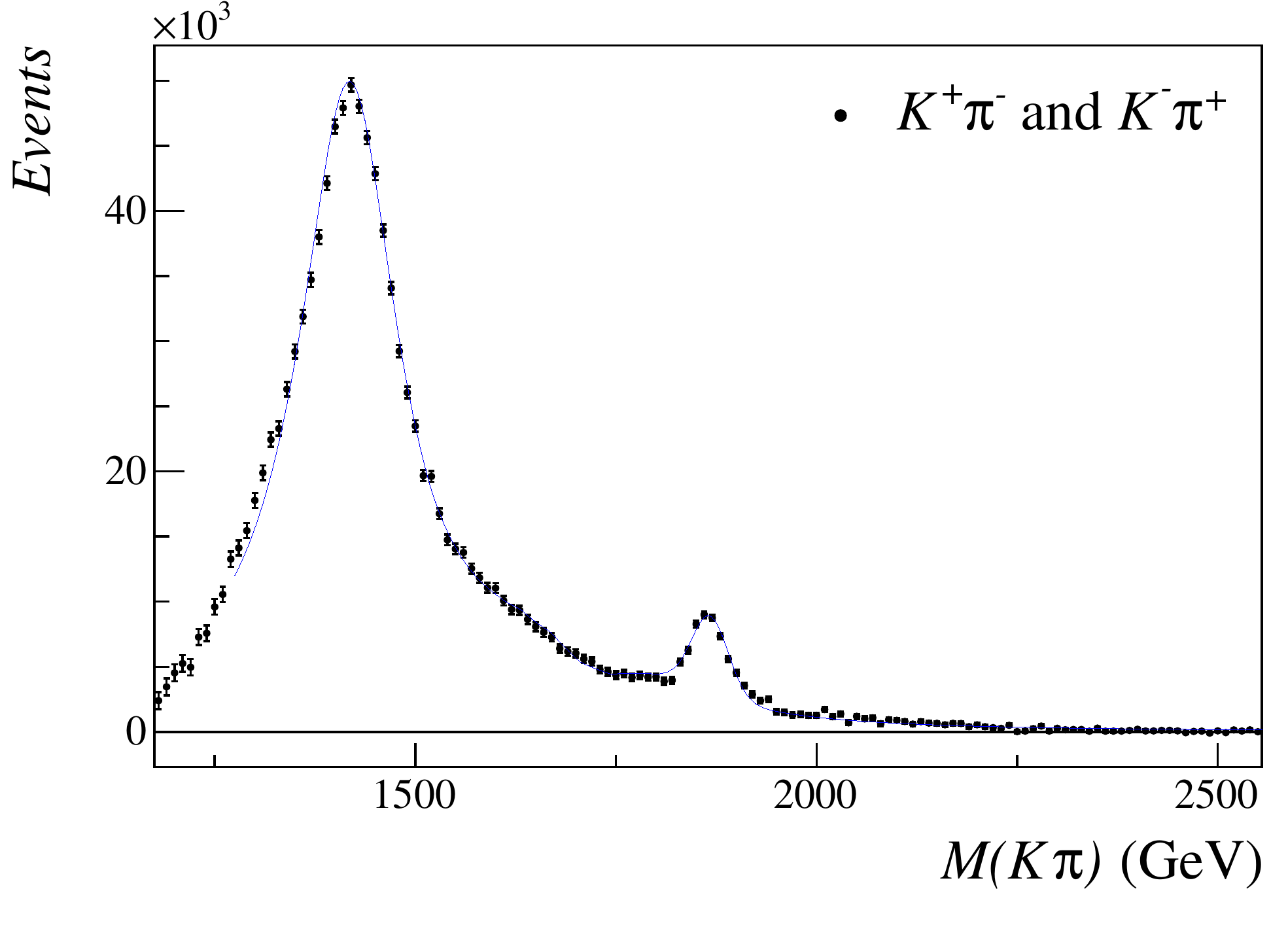}{75}{40}{(b)}
\OPicTwo{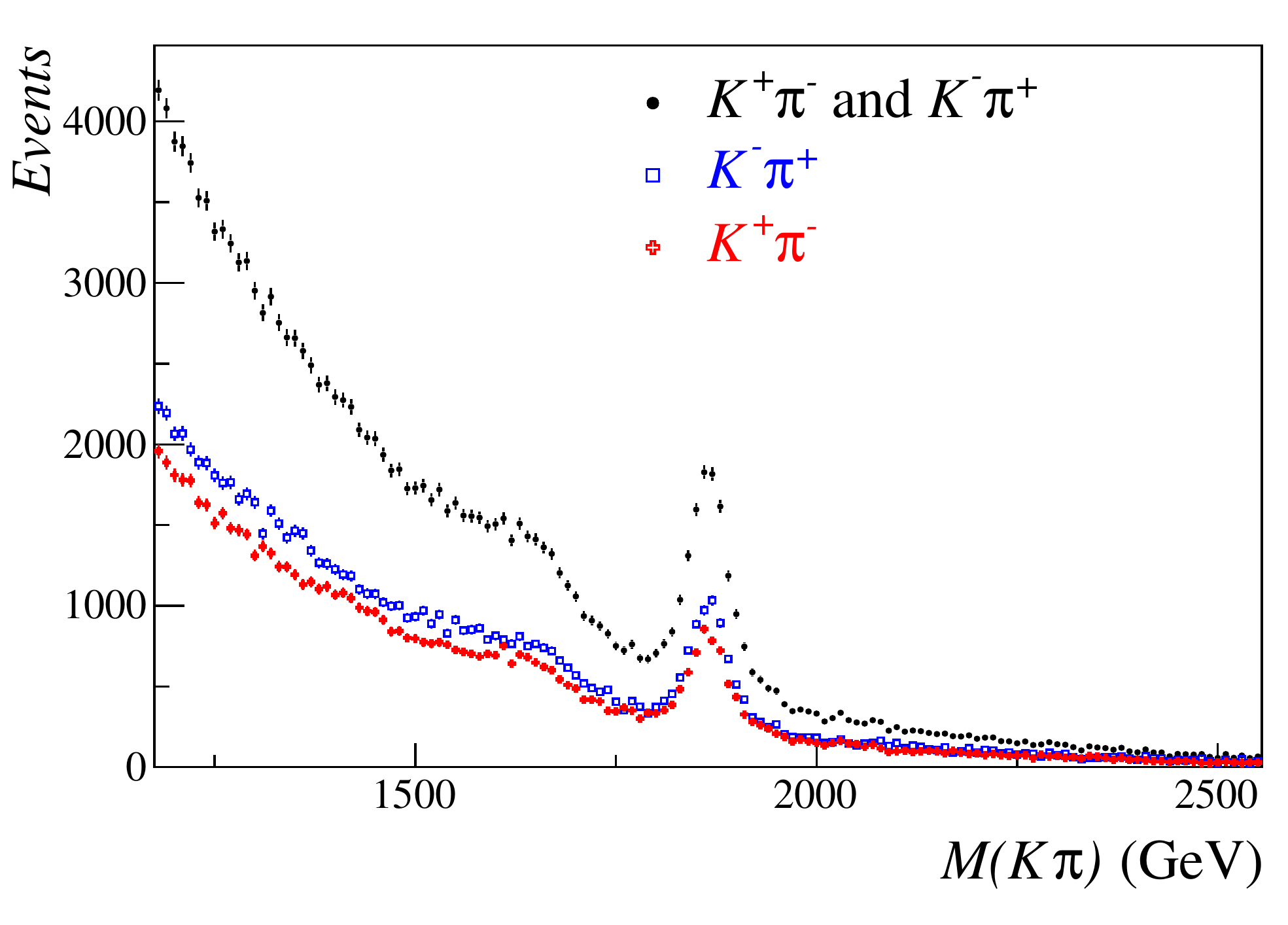}{75}{40}{(c)}
\OPicTwo{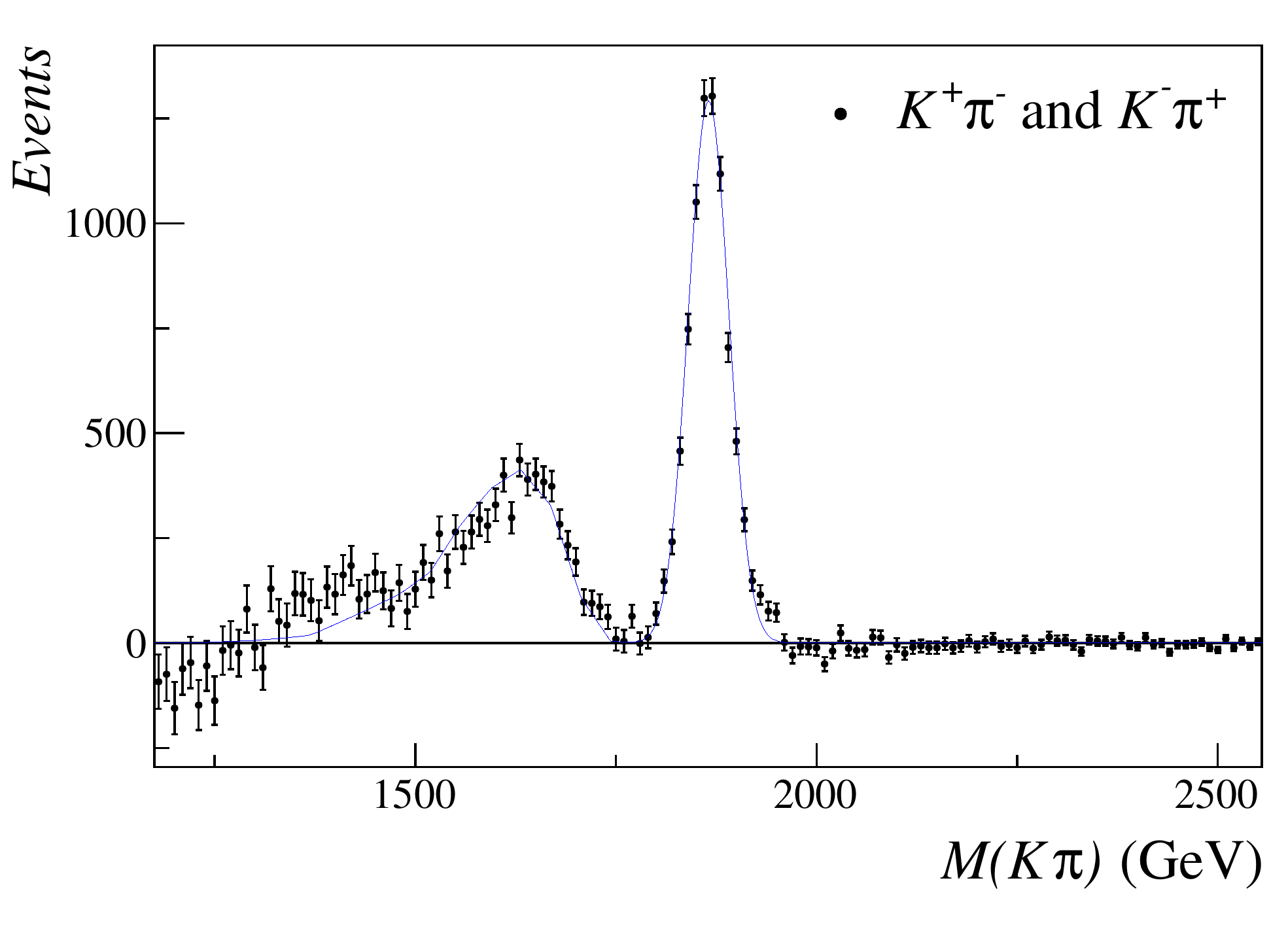}{75}{40}{(d)}
\caption{Invariant $ M(K\pi)$ mass spectra within a window of $\pm$700 \mmass\
around the nominal $\DZ$ mass. (a) $\DZ$ sample before and (b) after background
subtraction, (c) $\DS$ sample before and (d) after background subtraction. Both
neutral charge combinations are shown separately, together with their sum in (a)
and (c). See text for the background subtraction by fits.
\label{fig:ALL-DsD0_mass_all_charges}}
\end{figure}

The invariant $\Kpi$ mass spectra are shown in
Fig.~\ref{fig:ALL-DsD0_mass_all_charges}a for the untagged $\DZ$ sample, for all
neutral $\Kpi$ combinations and also separately for the $\Kpipm$ and $\Kpimp$
combinations. These spectra exhibit the $\DZ$ peak at 1865 \mmass. The
prominent peak to the left is due to the decay of the narrow $K_2^*(1430)$. In
Fig.~\ref{fig:ALL-DsD0_mass_all_charges}c invariant mass spectra are shown for
the tagged $\DZ$ sample. In this case, only some feed-through of the
$K_2^*(1430)$ resonance is seen and a pronounced, rather narrow peak about 250
\mmass\ below the nominal $\DZ$ mass. As shown by Monte Carlo simulations, this peak
at 1620 \mmass\ results from 3-body decays of the $\DZ \rightarrow K\pi\pi^0$,
where the $\pi^0$ escaped detection, with some contributions from $\DS$ decays
with more than 3 particles in the final state. The signal-over-background (S/B) ratio
is about 1:1 for the events of the tagged $\DZ$ sample. For the untagged
$\DZ$ sample, $S/B$ is only 1:10, but the number of signal events is four times
higher\footnote{The S/B is calculated in a $\pm 50$ \mmass\ mass window ($\pm 2
  \sigma)$ around the $\DZ$ peak.}.

\begin{figure}
\includegraphics[width=75ex]{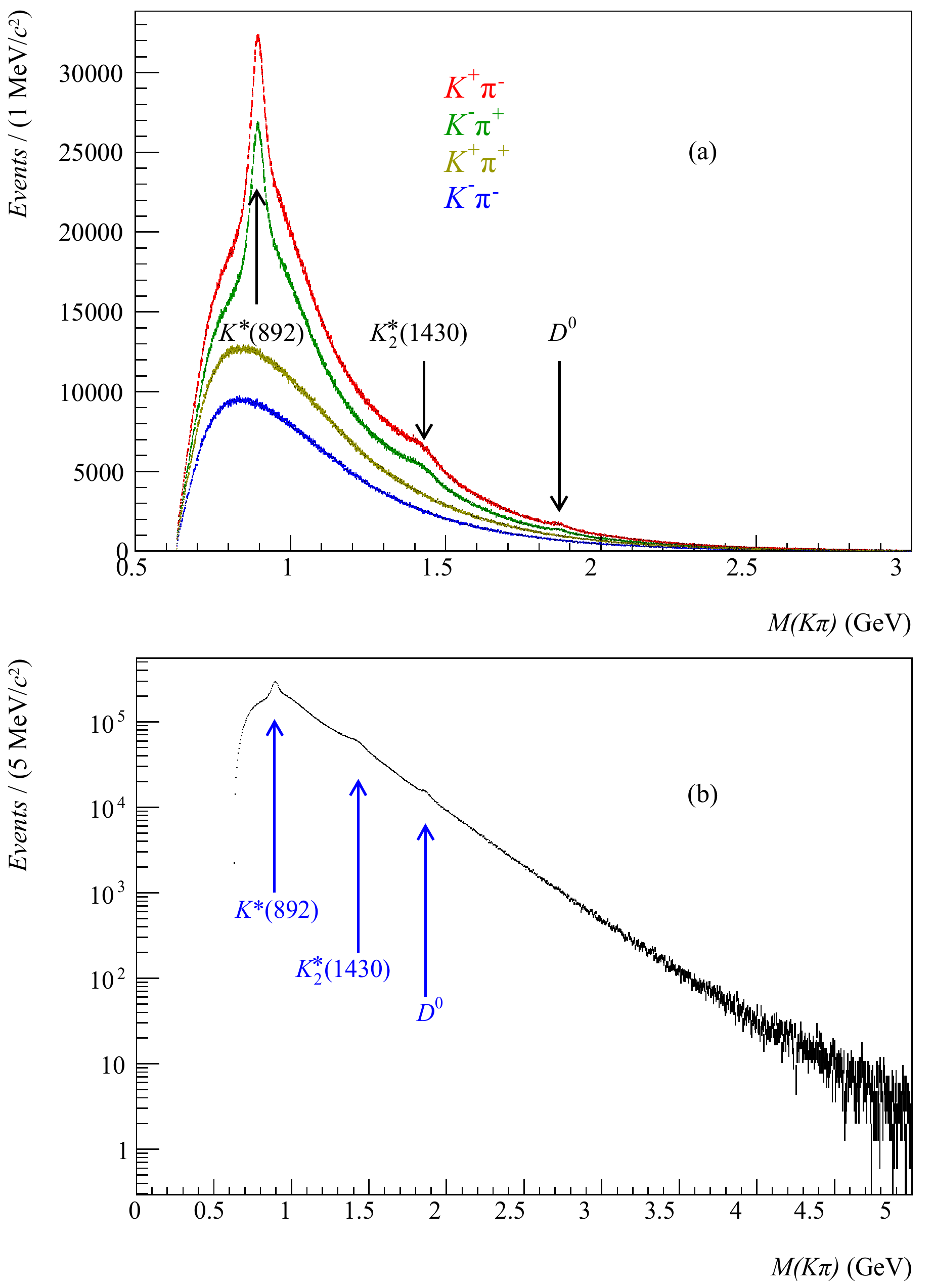}
%\OPicOne{Kpi-all-charges.pdf}{75}{50}{(a)}
%\OPicOne{Kpi2.pdf}{75}{50}{(b)}
%\OPicOne{fig4a.pdf}{75}{50}{(a)}
%\OPicOne{fig4b.pdf}{75}{50}{(b)}
\caption{ $\Kpi$ invariant mass distribution for the full mass range using data
from 2004 only. (a) All four charge combinations (from top to bottom + --, -- +,
-- --, + +) are shown separately, with a linear vertical scale; (b) only neutral,
non-exotic charge combinations (-- +, + --) summed up, logarithmic vertical
scale. \label{fig:Kpi2006all}}
\centering
\end{figure}

Mass spectra of all the $\Kpi$ combinations are shown in
Fig.~\ref{fig:Kpi2006all}a separately, using only data from 2004. The spectra
for the two neutral charge combinations show three narrow peaks corresponding to
$K^*(890)$, $K_2^*(1430)$ and $\DZ (1865)$. Also, other short lived kaonic
(strange) resonances are present but they superimpose together with
combinatorial background to a structureless distribution that can almost
perfectly be described by a single exponential function, see
Fig.\ref{fig:Kpi2006all}b.

\section{Method of signal extraction}
\label{sec:signal}
The invariant $K \pi$ mass spectra shown in
Figs.~\ref{fig:ALL-DsD0_mass_all_charges}a,c are fitted with a function given by
the sum of the following elements: a Gaussian for the $\DZ \ra K \pi $ signal,
an exponential for the background, a shape determined by Monte Carlo simulations
for the peak at 1620~\mmass\ from 3-body decays of the $\DZ$ and by relativistic
Breit-Wigner intensities for the $\KTS$ and $\KThS$. The latter $K$ resonance is
barely visible in this spectrum but shows up clearly in certain kinematic
regions, see below.

The $K\pi$ spectra are remarkably well described fitting them with 12
parameters, as described above.  Figs.~\ref{fig:ALL-DsD0_mass_all_charges}b,d
show the spectra after subtraction of the exponential background. From the fits
one obtains (3610 $\pm$ 90) $\DSp \ra (K^- \pi^+) \pi^+_s$ and (4530 $\pm$ 100)
$\DSm \ra (K^+ \pi^-) \pi^-_s$ for the tagged sample as well as (15200 $\pm$
800) $\DZ \ra K^- \pi^+$ and (18400 $\pm$ 900) $\barDZ \ra K^+ \pi^-$ for the
untagged $\DZ$ sample.

The dependence on kinematic variables of the production rate of $\DZ$ and $\DS$,
together with those of the neighbouring $\KTS$ resonance and the background, is
extracted by fitting the mass spectra for each kinematic bin. Alternatively, the
signal distributions of the $\DZ$ and the $\DS$ are obtained by side-band
subtraction.

\begin{figure}
\centering
\includegraphics[width=\textwidth]{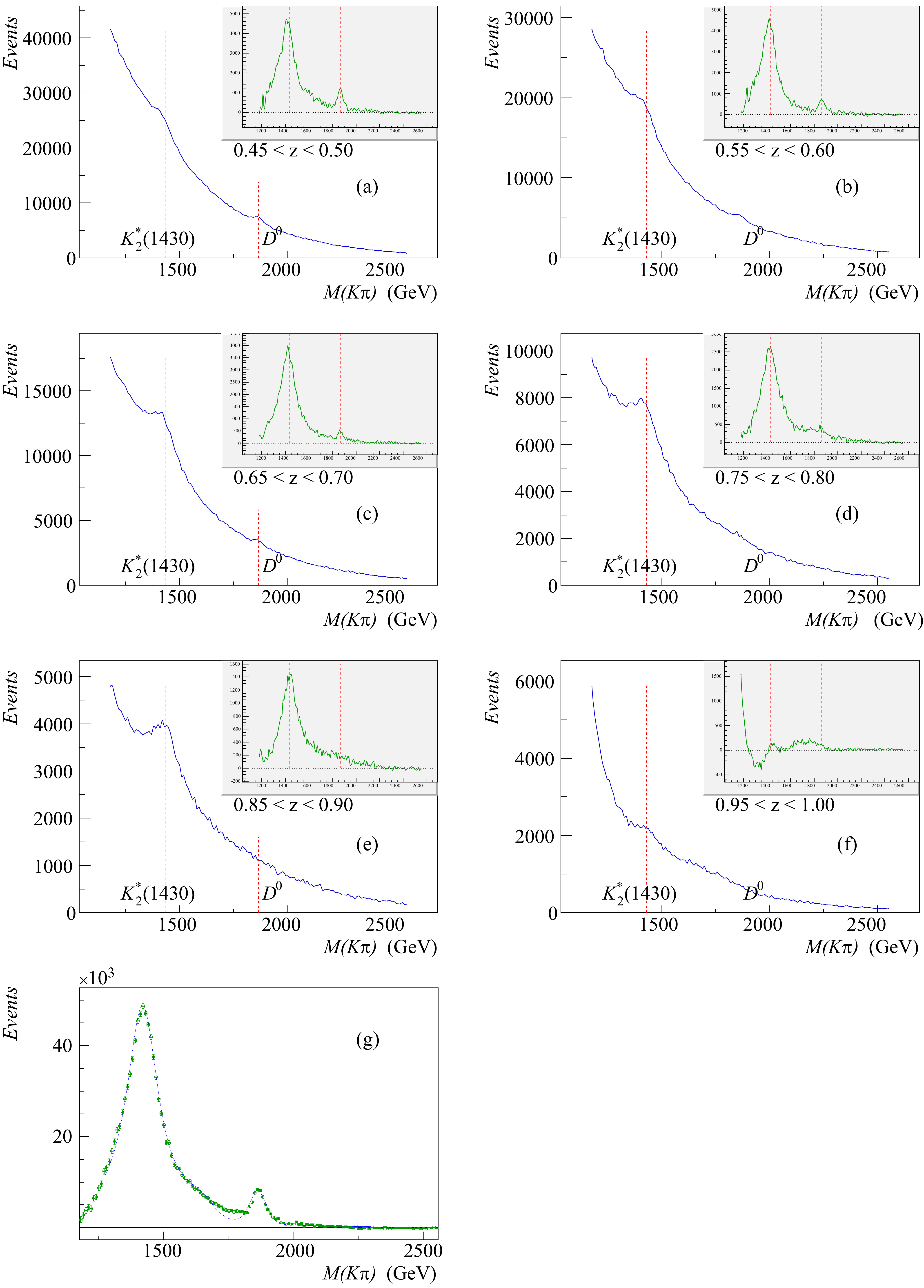}
\caption{a--f) Invariant $K \pi$ mass spectra in bins of the energy fraction $z$ for
the untagged $\DZ$ sample. The vertical (red) lines indicate the nominal
positions of $\KTS$ and $\DZ$. The inserts demonstrate the signal
behaviour after the removal of the fitted exponential background. The fit
contains $\DZ\rightarrow K\pi$ at 1865~\mmass, $\DZ\rightarrow K\pi \pi^0$ at
1620~\mmass, the $\KTS$ and an exponential background. g) shows the signal
behaviour after removal of the fitted exponential background for the entire $z$ 
range (no $\KThS$  assumed). This figure has to be compared with
Fig.~\protect\ref{fig:ALL-DsD0_mass_all_charges}b where the $\KThS$  was included
in the fit.\label{fig:D0:resonance_z}}   
\end{figure}

Using the first method, the fit yields in every bin of a given kinematic
variable the number of $D$, $\KTS$ and $\KThS$ together with the background. In
Figs.~\ref{fig:D0:resonance_z}a-f examples of the $K\pi$ invariant mass spectra
for different intervals in $z$ are shown before and after the subtraction of the
fitted exponential background. These fits did not include the $\KThS$. The
fitting method allows monitoring of all details of the fit, as illustrated in
the inserts of Fig.~\ref{fig:D0:resonance_z}. The broad structure showing up for
$z>0.75$ in Fig.~\ref{fig:D0:resonance_z} is attributed to the $\KThS$
resonance.  This resonance follows the same behaviour as the $\KTS$ resonance,
i.e.\ it is produced at larger values of $z$ than the $\DZ$ (see
sec.~\ref{kinedis}). The introduction of the $\KThS$ resonance in the final fit
also removed a small but statistically significant and unexplained discrepancy
between fit and data on the left side of the $\DZ \ra K \pi$ peak in the
$z$-integrated spectrum, where the fit before the inclusion of the $\KThS$ was
systematically below the data (see Fig.~\ref{fig:D0:resonance_z}g to be compared
with Fig.~\ref{fig:ALL-DsD0_mass_all_charges}b, where the $\KThS$ has been
included).

The second method for signal extraction is the standard side-band subtraction.
This method can only be applied to the $\DZ$ and the $\DS$ signals, due to the
limited mass range ($\pm 700$ MeV around the nominal $\DZ$ mass). Three $\Kpi$
mass windows are chosen. The central one, which is 100 \mmass\ wide and centered
at the nominal $\DZ$ mass, contains the $\DZ \ra \Kpi$ signal plus background.
The two side-bands contain only background events. They are 50 \mmass\ wide and
centered at $\pm$100 \mmass\ above or below the nominal $\DZ$ mass. Thus three
independent distributions are obtained as a function of each kinematic
variable. The sum of the side-band distributions is subtracted from the central
distribution, assuming that the side-band distributions correctly represent the
distribution of background under the signal. This assumption is supported by the
observed similar behaviour of the distributions in the two side-bands.

Usually, the background below the signal is obtained by linear interpolation
between the side-bands. Such a linear interpolation overestimates the
background under the signal. Therefore it cannot be applied for the untagged
$\DZ$ sample, where $S/B\sim 1/10$. Instead, an estimate of the background under
the signal is obtained from the fit. The total number of background events in
the two side-bands is correspondingly rescaled.

For the chosen width of the central window, about 5\% of the signal is found
outside. Hence the number of signal events  obtained by side-band subtraction is
expected to be lower by 5\% than that obtained with the signal fitting method.

\section{Comparison of kinematic distributions}
\label{kinedis}

In this section, event distributions are shown as a function of the relevant
kinematic parameters, for both the tagged and untagged $\DZ$ samples as well as
for the $\KTS$ and background. The data collected in 2002-2006 are used, and the
distributions are not corrected for acceptance. However, the geometric
acceptances for the various compared $\Kpi$ systems are similar.

The distributions of the $\KTS$ signal are obtained from the untagged $\DZ$
sample using the signal fitting method. The distributions of the
$\Kpi$-background combinations are extracted from the two $\Kpi$ side bands of
the tagged $\DZ$ sample, at invariant masses of $1765 \pm 50$ and $1965 \pm 50$
\mmass. The kinematic distributions of $\DZ$ and $\DS$ are obtained by applying
both signal extraction methods described above, allowing to cross-check the
stability of the result. While for the tagged $\DZ$ sample perfect agreement is
found between the two methods, for the untagged sample some disagreement beyond
the statistical error is observed, for instance at low values of $z$ or low
$\Kpi$ energy $E$. This is the result of strongly varying background shapes with
additional broad resonances emerging below the $\KTS$. The corresponding data
points for $\DZ$ and $\KTS$ are omitted, since a more complex background
description would be needed.

\begin{figure}
\centering
\OPicTwo{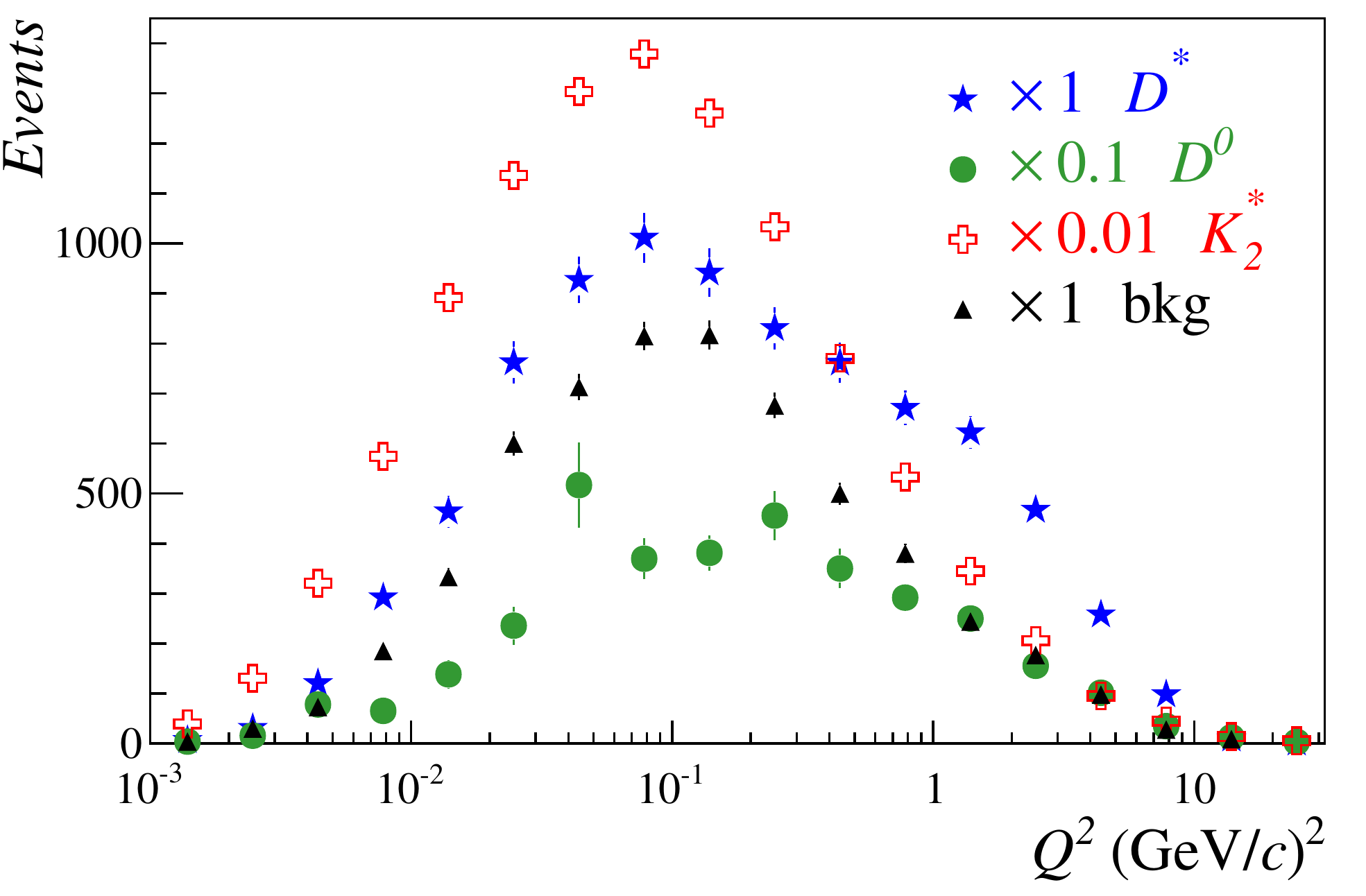}{18}{57}{(a)} \hfill
\OPicTwo{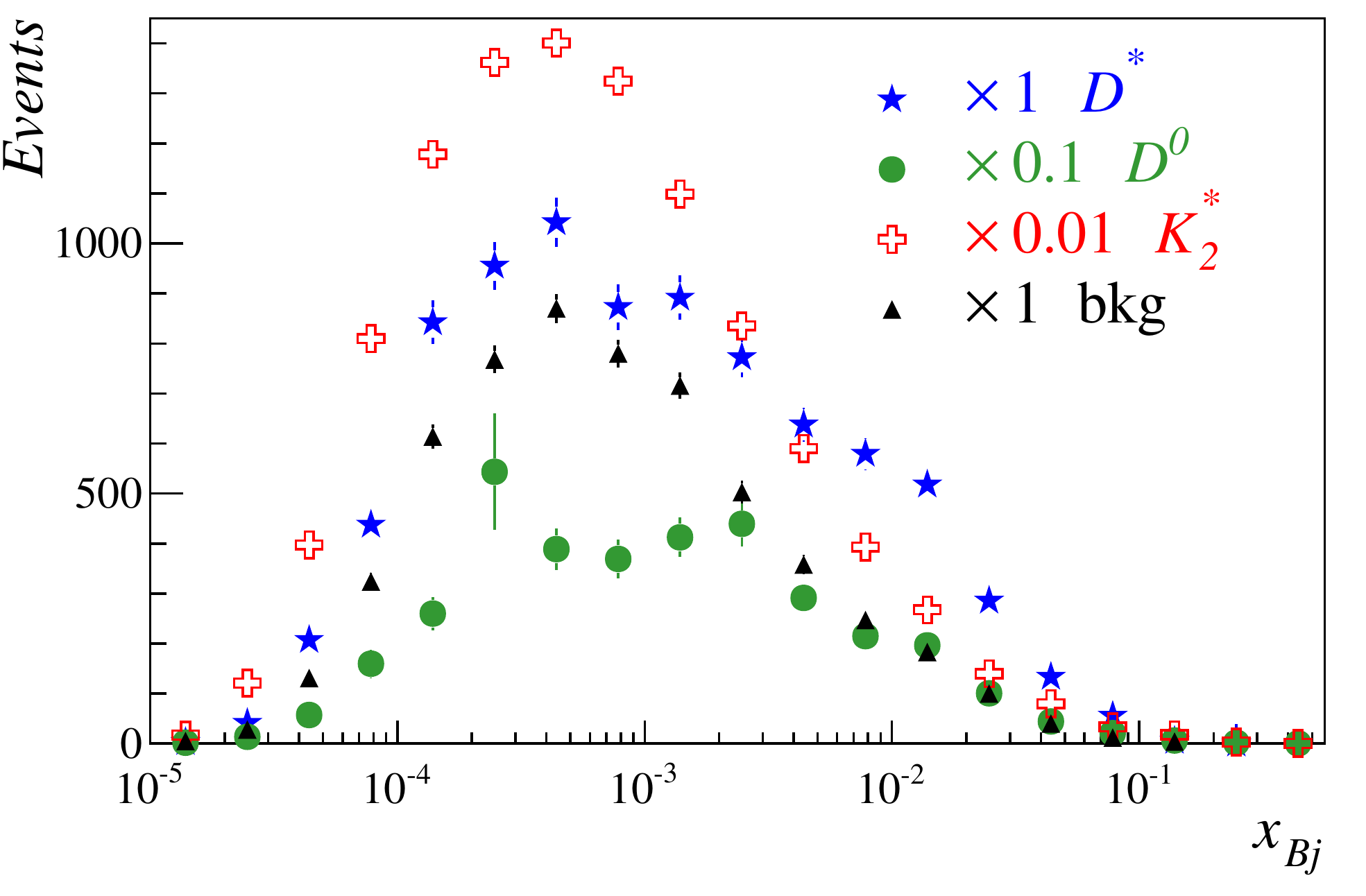}{18}{57}{(b)} \\
\OPicTwo{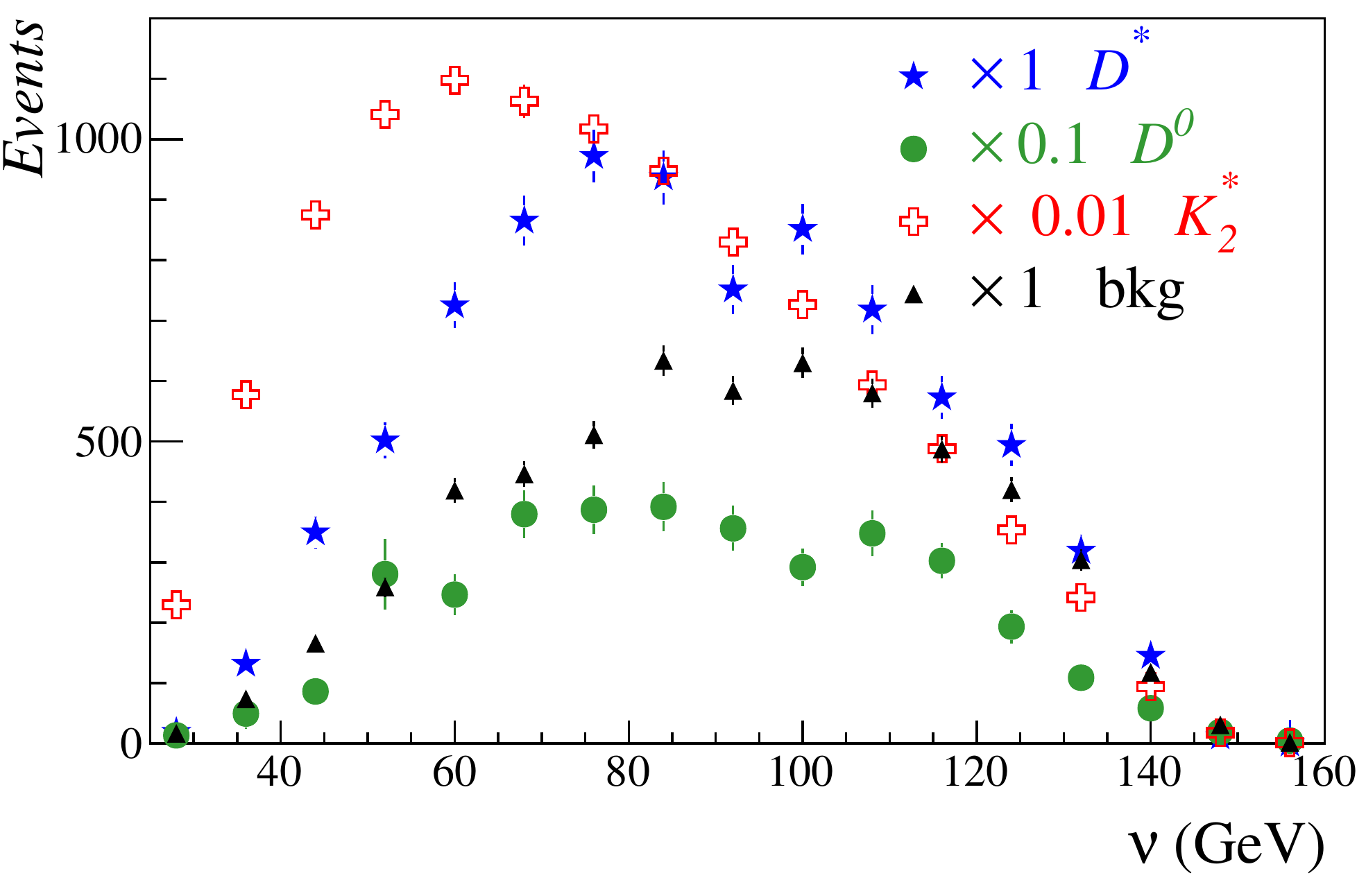}{18}{57}{(c)} \hfill
\OPicTwo{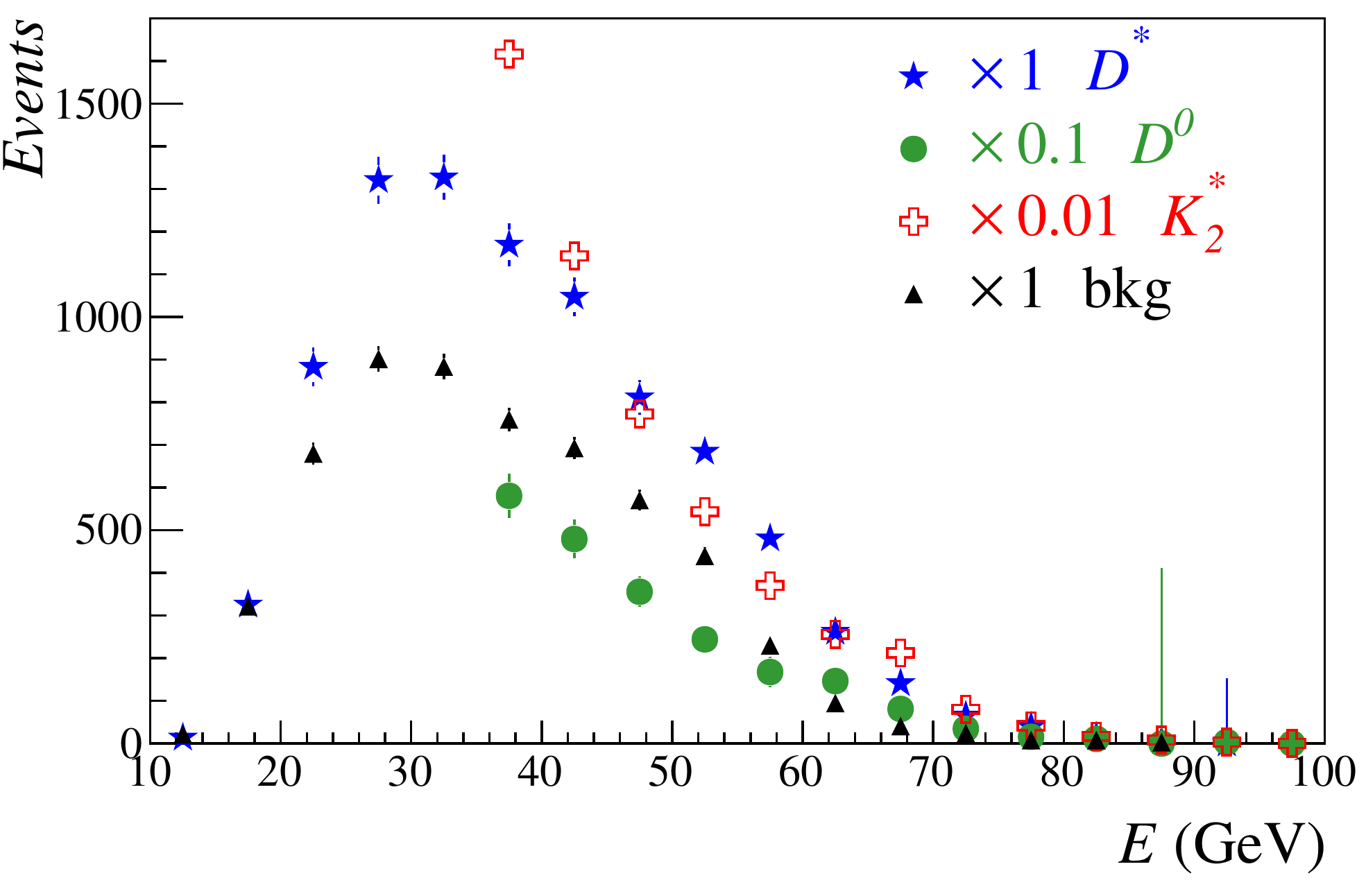}{18}{57}{(d)} \\
\OPicTwo{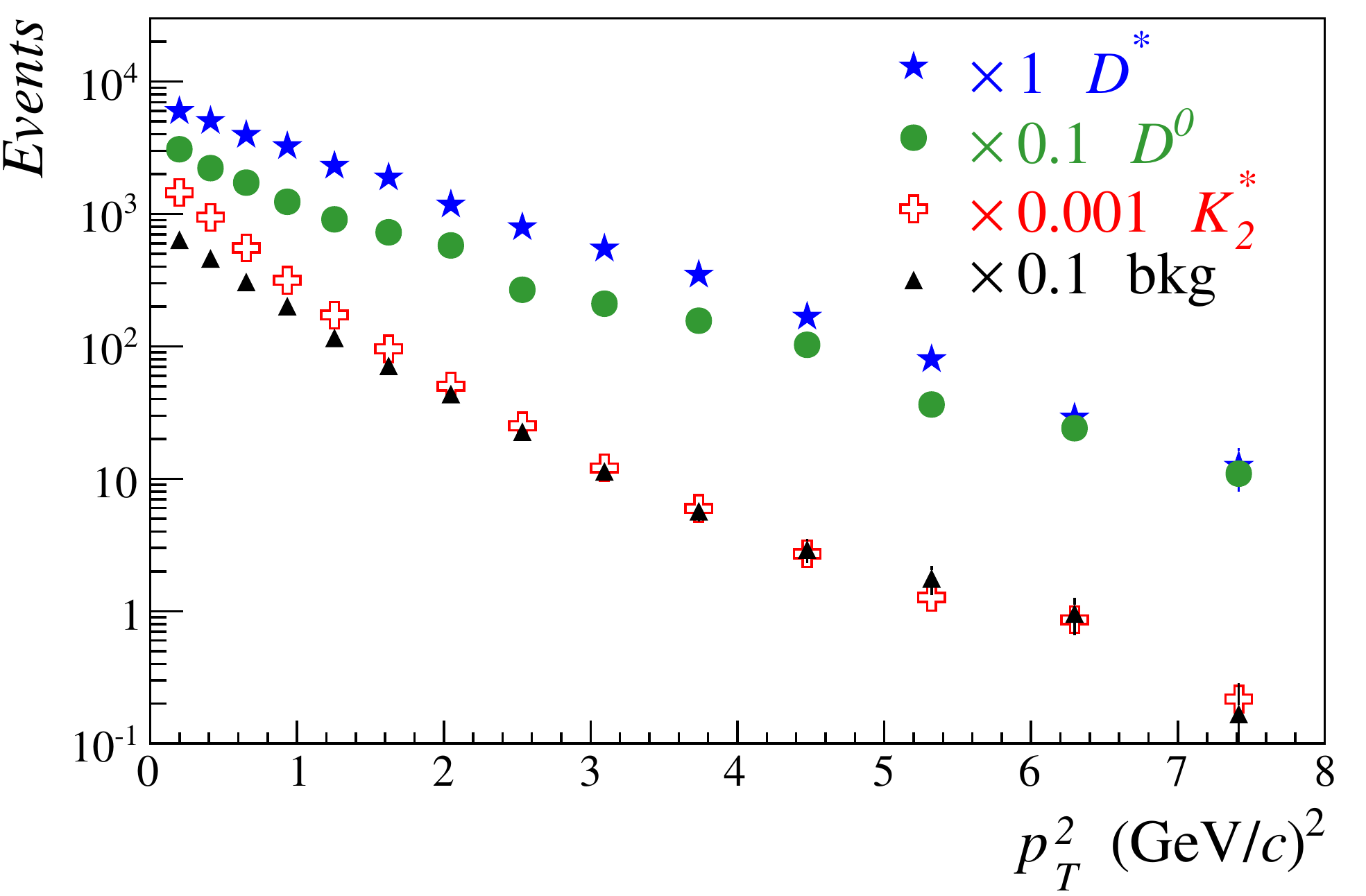}{18}{18}{(e)} \hfill
\OPicTwo{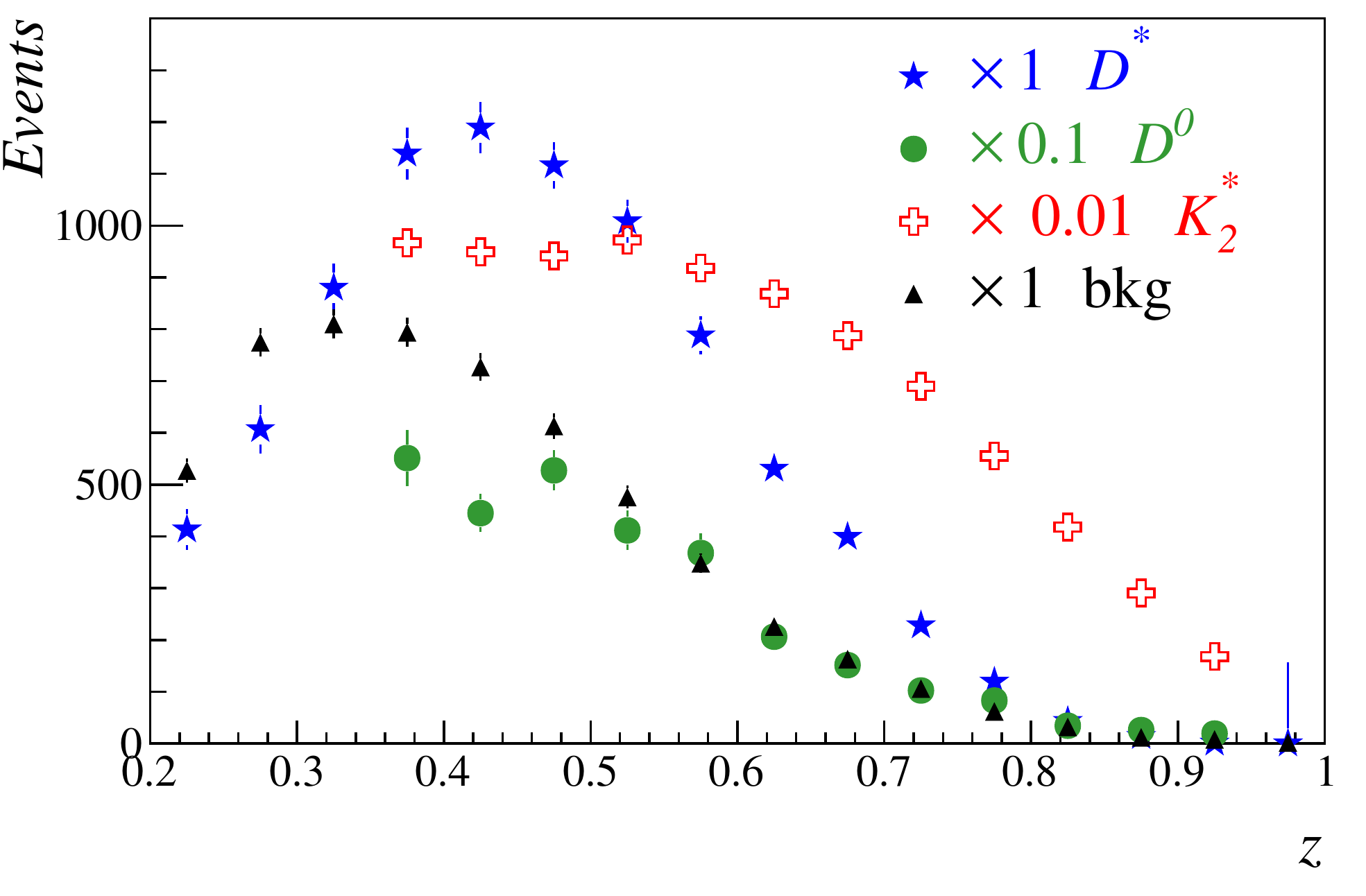}{18}{18}{(f)}
\caption{Measured kinematic distributions of various ($\Kpi$) systems before
acceptance correction as a function of 
(a) $Q^2$,  
(b) $x_{Bj}$, 
(c) $\nu $, 
(d) $E$, 
(e) $p_T^2$, and 
(f) $z$. 
The symbols $\DZ$ and  $K^*_2$ denote $\DZ$ and $\KTS$
from the untagged sample. The symbols $\DS$ and bkg denote $\DS$ and background
from the two side-band windows for the $\DS$ sample. 
The data are from the years 2002 to 2006. 
\label{fig:Ds/ALL-Q2}}
\end{figure}

In Fig.~\ref{fig:Ds/ALL-Q2}, the distributions of the $\DZ$, the $\KTS$ and the
background under the $\DZ$ are compared, showing their different behaviour. The
distributions as a function of the inclusive variables $Q^2$ and $x_{Bj}$ are
displayed in Figs.~\ref{fig:Ds/ALL-Q2}a,b. For the tagged sample, the average
values of $Q^2$ and $x_{Bj}$ extracted from these distributions are about 0.5
\gomt\ and 0.005, respectively. Some differences between signal and background
events are observed at large values of $Q^2$ and $x_{Bj}$. As a function of $
\nu $, the distributions for the various $\Kpi$ systems are significantly
different (see Fig.~\ref{fig:Ds/ALL-Q2}c). The $\KTS$ distribution peaks at
lower values than that of $\DZ$, and the rise at low $\nu$ that is caused by the
increase of both acceptance and cross-section starts at lower $\nu$. The
background peaks at a somewhat higher values, but has a similar rise with $\nu$
as $\DZ$ and $\DS$.

No clear differences are observed between the distributions as a function of the
$\Kpi$ energy $E$ (see Fig.~\ref{fig:Ds/ALL-Q2}d). Given the reason described
above, the $\KTS$ and the $\DZ$ data points at lower values of $E$ are omitted
from the untagged sample.  The distribution for the $\DS$ signal as a function
of $p_T^2$ (Fig.~\ref{fig:Ds/ALL-Q2}e) shows an almost single-exponential
decrease, while the distribution for the $\DZ$ flattens above 3 \gomt. The
difference between $\DZ$ and $\DS$ may be related to the fact that for the $\DS$
only the $p_T^2$ of the 2-body subsystem is shown. Both distributions are
significantly different from those of background and $\KTS$. From a fit of an
exponential function up to $p_T^2 = 2 \gomt$, the following slopes are obtained
in units of \gommt: $-0.84\pm 0.03$ for $\DS$, $-0.96\pm 0.06$ for $\DZ$,
$-1.94\pm 0.01 $ for $\KTS$ and $-1.69\pm 0.01 $ for background. The
distributions in $z$ show significant differences, too. The background is
concentrated at smaller values of $ z$ than the $\DZ$ signal. Moreover, the
distribution of the $\KTS$ is peaked at significantly higher values of $z$ than
that of the $\DZ$.

In conclusion of the comparison: remarkable differences are observed between the
distributions of the $D$ meson signals, the $\KTS$, and the background as a
function of the kinematic variables $\nu$, $p_T$ and $z$. This clearly points
to different production mechanisms for $D$ mesons and the $\KTS$. The observed
differences between $D$ mesons and $\KTS$ agree qualitatively with the
differences expected if the $D$ mesons result from the fragmentation of a pair
of charm quarks and the $\KTS$ from the fragmentation of a quark knocked out in
a leading order process.

The interpretation of the kinematic distributions of the background is more
complex, since this background is dominated by combinatorial entries. No attempt is
made to interpret it. However, one should mention that other background
events of non-combinatorial origin (e.g.\ in the untagged sample the background
taken from side bands has also large contributions from resonances or from $\pi
K$ correlated  production in the fragmentation) have been observed to behave
very similar to the background shown in Fig.~\ref{fig:Ds/ALL-Q2}. 

%The interpretation of the background is more complex, since it is dominated by
%combinatorial entries. For completeness one should mention that background
%events of non-combinatorial origin (e.g.\ in the untagged sample the background
%taken from side bands has also large contributions from resonances or from $\pi
%K$ correlated  production in the fragmentation) behave very 
%similar to the background shown in Fig.~\ref{fig:Ds/ALL-Q2}.
\section{Acceptance and integrated luminosity}
\label{sec:acceptance}

Acceptances and integrated luminosity, which are needed to extract
semi-inclusive total and differential cross-sections, are calculated only for
the tagged $\DZ$ sample of the year 2004. Since this is the first detailed
acceptance calculation for this particular final state at COMPASS, the present section also
aims at illustrating the acceptances of the COMPASS spectrometer for the
detection of the scattered muon and the $\DSpm$. For this reason,
2-dimensional acceptances will be shown as a function of selected variables.

Acceptance calculations are done using a complete Monte Carlo simulation of the
detector configuration, including the triggers and the track reconstruction code
for the 2004 data. Events are generated using AROMA 2.2.4~\cite{AROMA}, which
assumes photon-gluon fusion into $\ccbar$ to be the dominant underlying
mechanism for $\DS$ production. Default fragmentation functions are used and
parton showers are generated. The charm quark mass is set to
1.35~\gmass. Produced $\DS$s are forced to decay to $\DZ \pi ^+ \ra K^- \pi ^+
\pi^+$ for $\DSp$ or to $\barDZ \pi ^- \ra K^+ \pi ^- \pi ^-$ for $\DSm$.
Trigger conditions and data selection criteria applied to the Monte Carlo events
are the same as for real data. In total, $10^7$ events were generated for both
decays. The acceptances are calculated as a function of the reconstructed values
of the kinematic variables, thus accounting for experimental resolution and
bin-to-bin smearing.

\begin{figure}
\centering
\OPicTwo{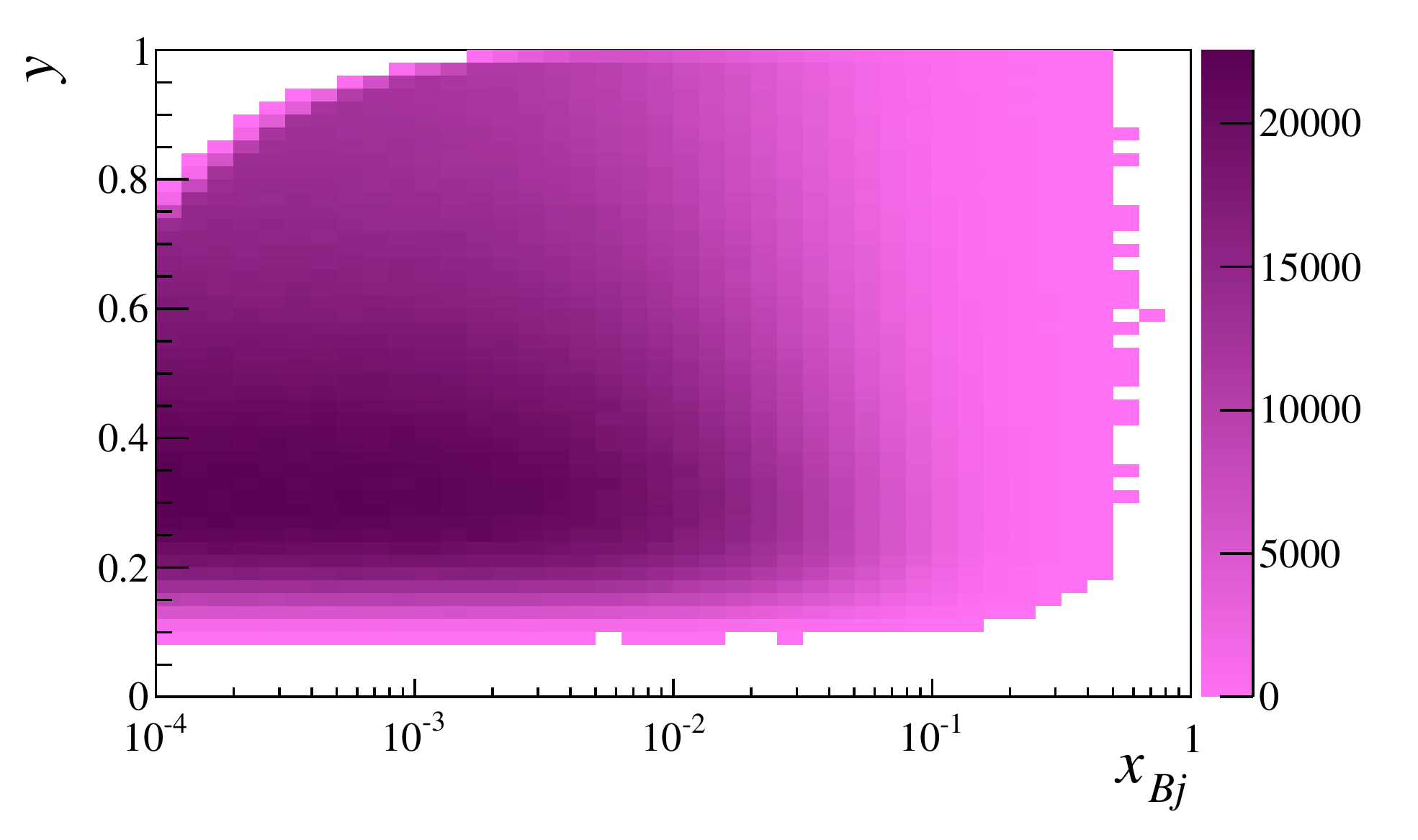}{70}{49}{(a)}
\OPicTwo{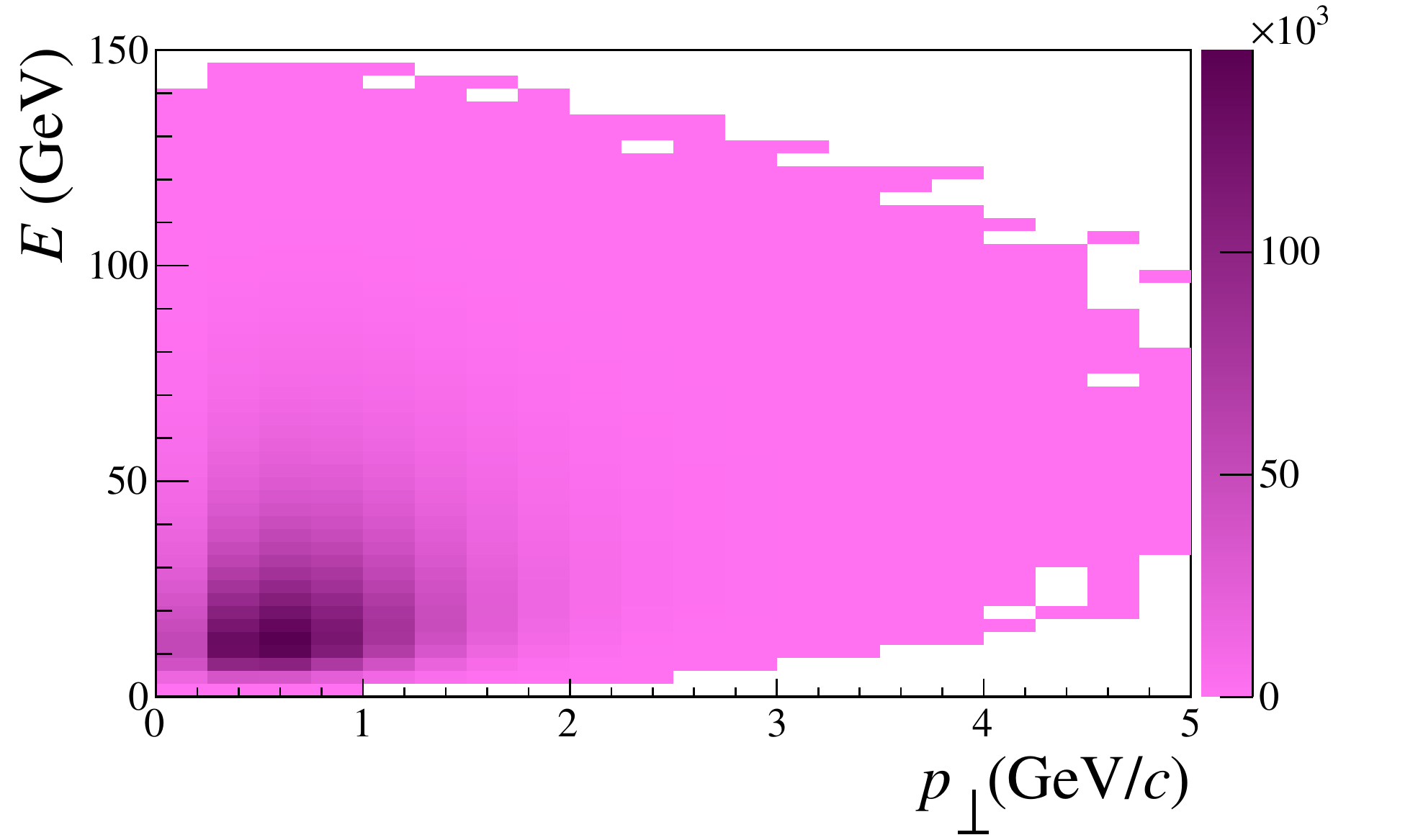}{70}{49}{(b)}
\caption{ 
Number of generated (AROMA) events as a function of (a) $x_{Bj}$ and $ y$
and (b) $p_\perp$ and $E$. \label{fig:MC:Muon acceptance} }
\end{figure}

Figure~\ref{fig:MC:Muon acceptance} shows the number of generated events (a) as
a function of $x_{Bj}$, $y$ and (b) as a function of $p_\perp$, $E$ of the D
meson. The transverse momentum $p_\perp$ is measured with respect to the
direction of the incoming muon beam. In both pictures the generated events are
mainly concentrated in the lower left corner.

\begin{figure}
\centering
\OPicTwo{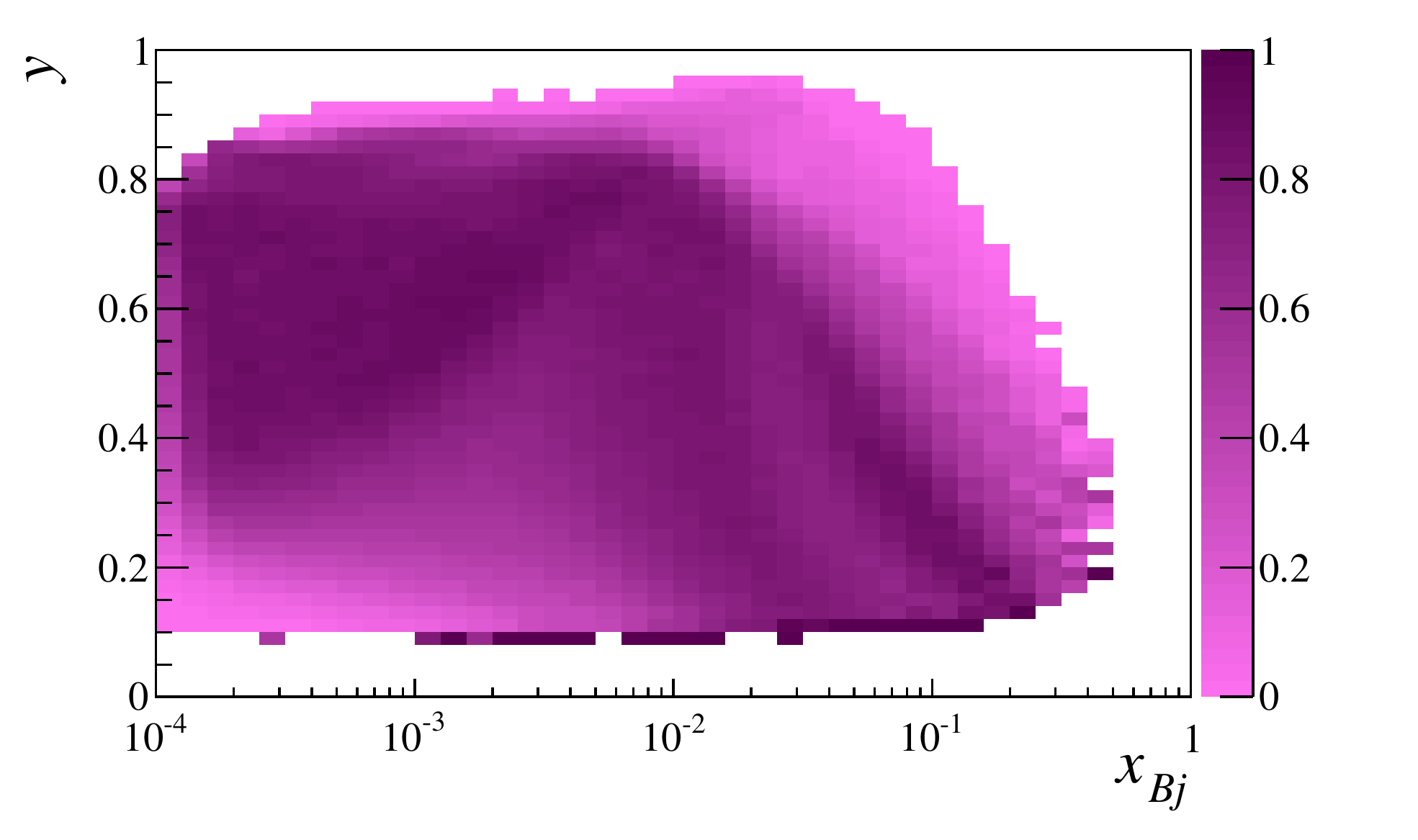}{70}{50}{(a)}
\OPicTwo{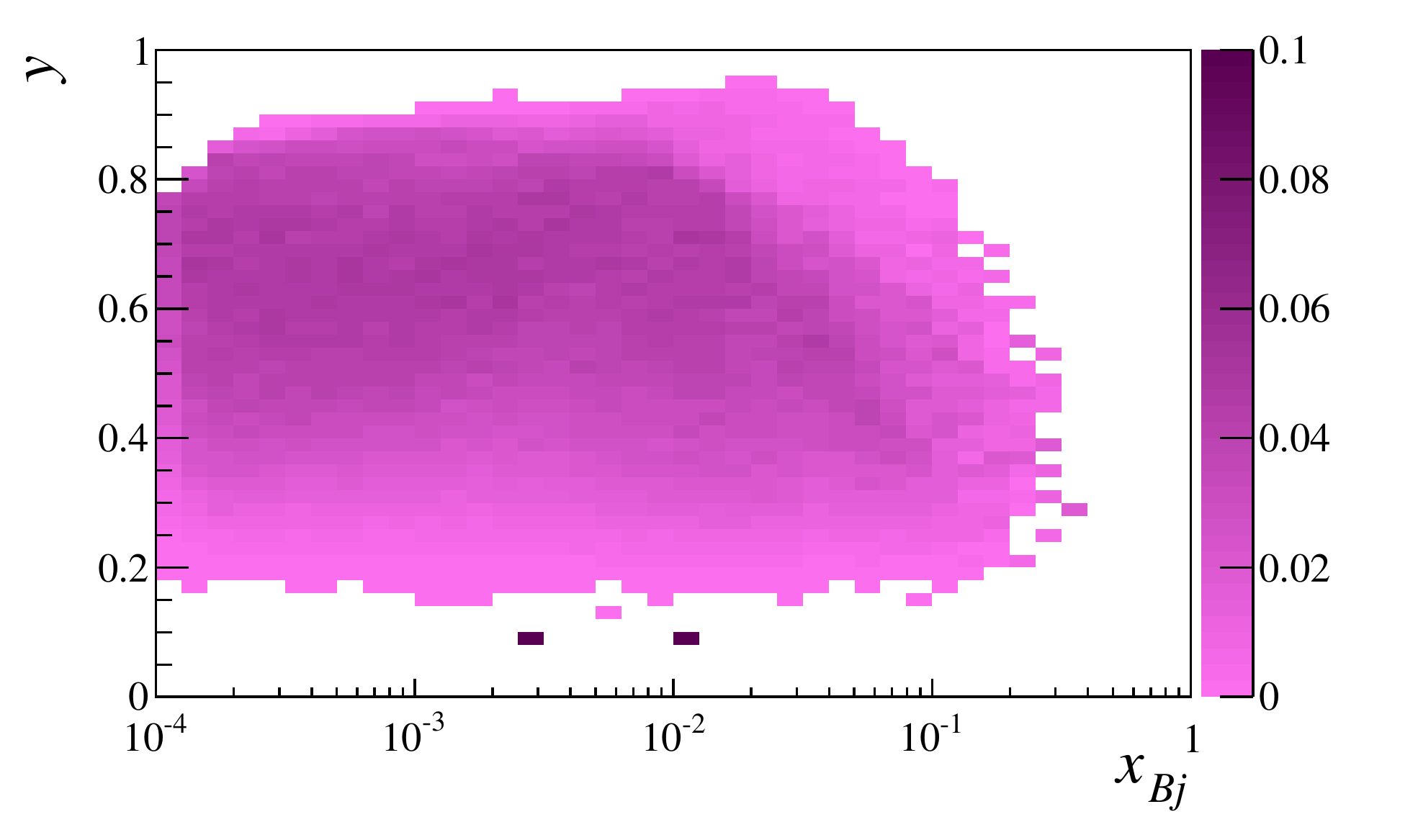}{70}{50}{(b)}
\caption{
(a) 'Inclusive' acceptance $ A_\mu (x_{Bj},y)$ and
(b) overall acceptance $A_{\DS}(x_{Bj},y)$ before applying the $E$
  cut. \label{fig:MC:Ds acceptance}}  
\end{figure}

For illustration, the acceptance for $\DS$ production is shown at two stages,
i.e.\ after requiring the reconstruction of the scattered muon and after the
additional reconstruction of the three hadrons from the $\DS$ decay. The
`inclusive' acceptance $A_\mu (x_{Bj},y)$ is shown in Fig.~\ref{fig:MC:Ds
acceptance}a, and the overall acceptance $A_{\DS}(x_{Bj},y)$ in
Fig.~\ref{fig:MC:Ds acceptance}b. In the kinematic region relevant for charm
production, the inclusive acceptance $A_\mu (x_{Bj},y)$ is fairly homogeneous
and ranges between 50\% and 80\%. The overall acceptance $ A_{\DS}(x_{Bj},y)$ is
also homogeneous for $y >0.2$ and ranges from 1\% to 5\%. The cut-off at $y=
0.2$ is due to the momentum selection for the RICH identification.
 
\begin{figure}
\centering
%\PicOne{fig9.pdf}
\includegraphics[width=0.6\textwidth]{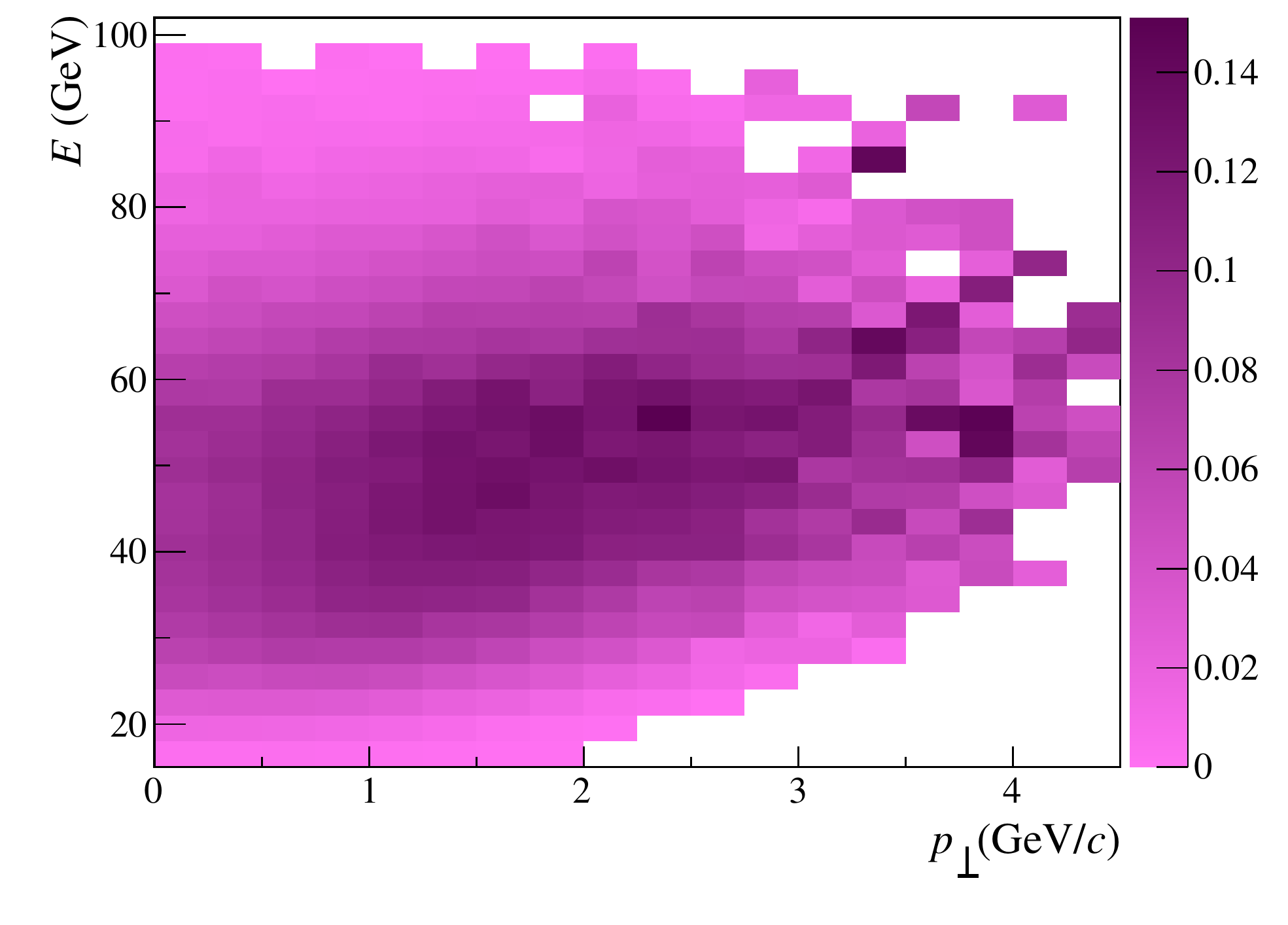}
\caption{Overall acceptance $A_{\DS}( p_\perp ,E)$ \label{fig:MC:acctable}}
\end{figure}

The overall acceptance $A_{\DS}$ as a function of $E$ and $ p_\perp$
(i.e.\ transverse momentum with respect the incoming muon) is shown in
Fig.~\ref{fig:MC:acctable}. The upper limit of about 100 mrad for the
spectrometer acceptance in the year 2004 can be seen at low energy and large
$p_\perp $. For $20\ \mrf{GeV} < E < 80\ \mrf{GeV}$ the acceptance ranges
between 5$\%$ and 13$\%$. Outside this energy region the acceptance drops to
zero due to the lack of particle identification and therefore $20\ \mrf{GeV} <E
< 80\ \mrf{GeV}$ is required in the further analysis. The one-dimensional
acceptances used below to determine the differential inclusive cross-sections
are limited to this range of $\DZ$ energies.

\begin{figure}
\centering
\OPicTwo{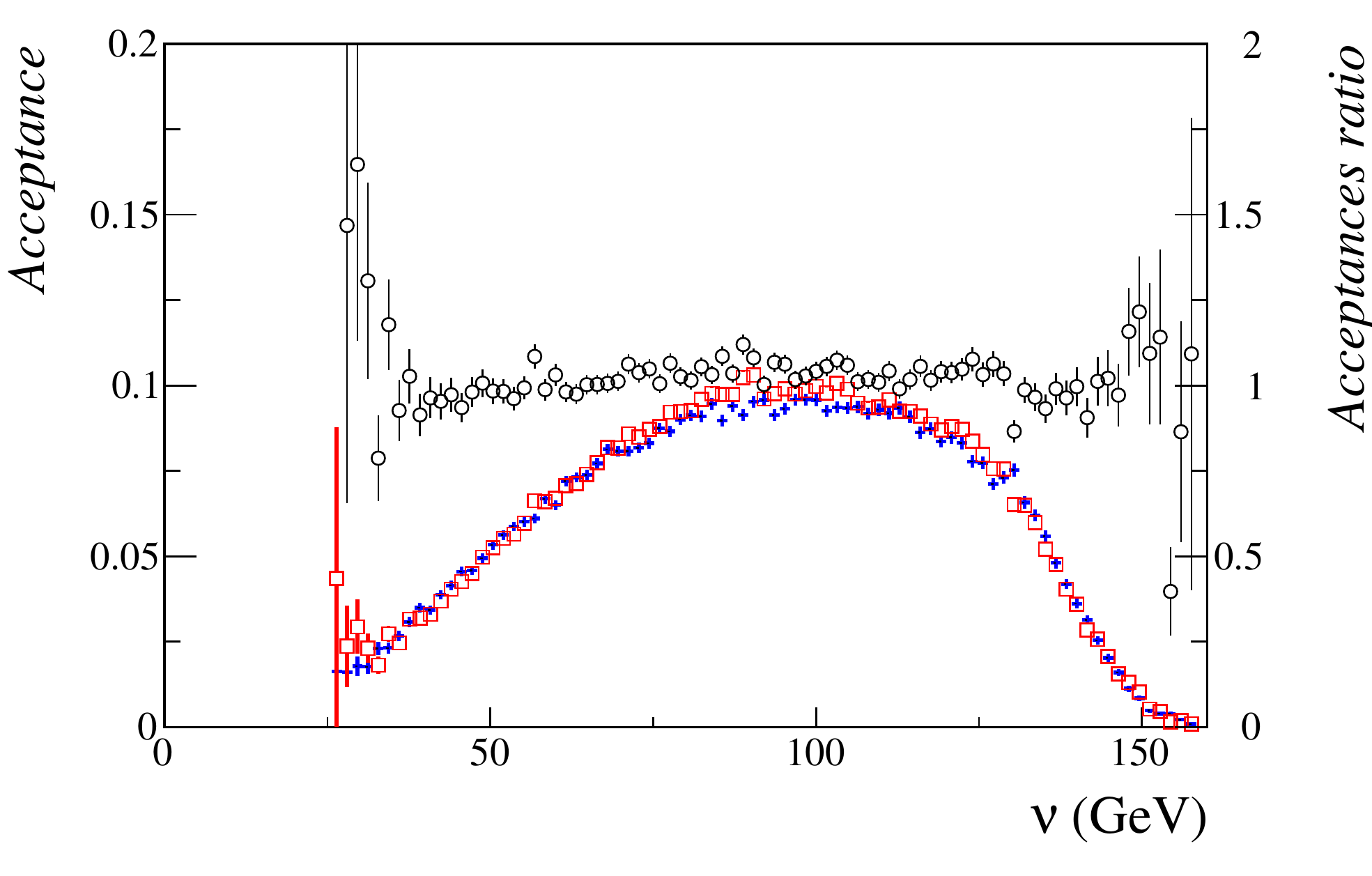}{35}{50}{(a)} \hfill
\OPicTwo{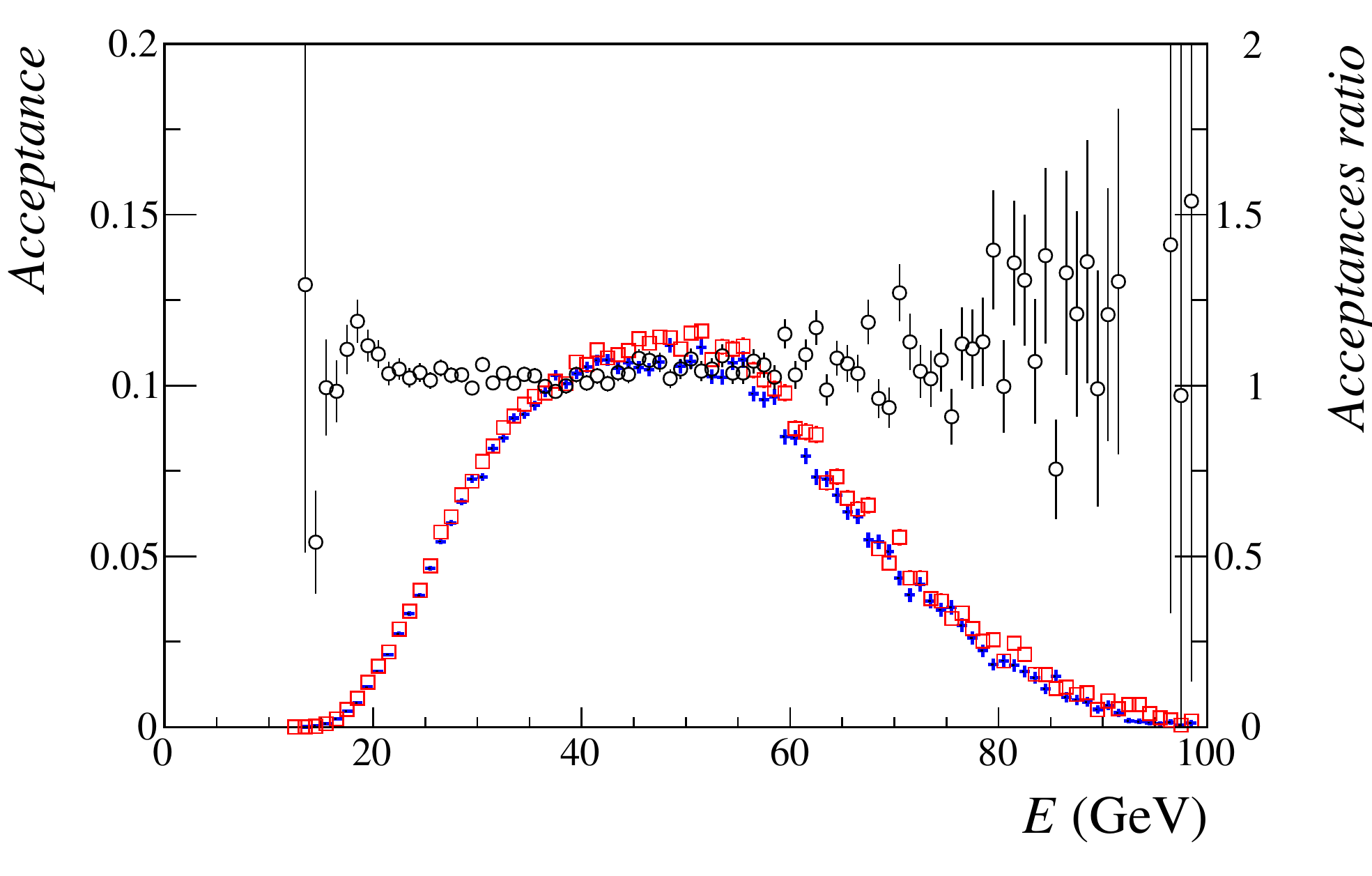}{35}{50}{(b)}  \\
\OPicTwo{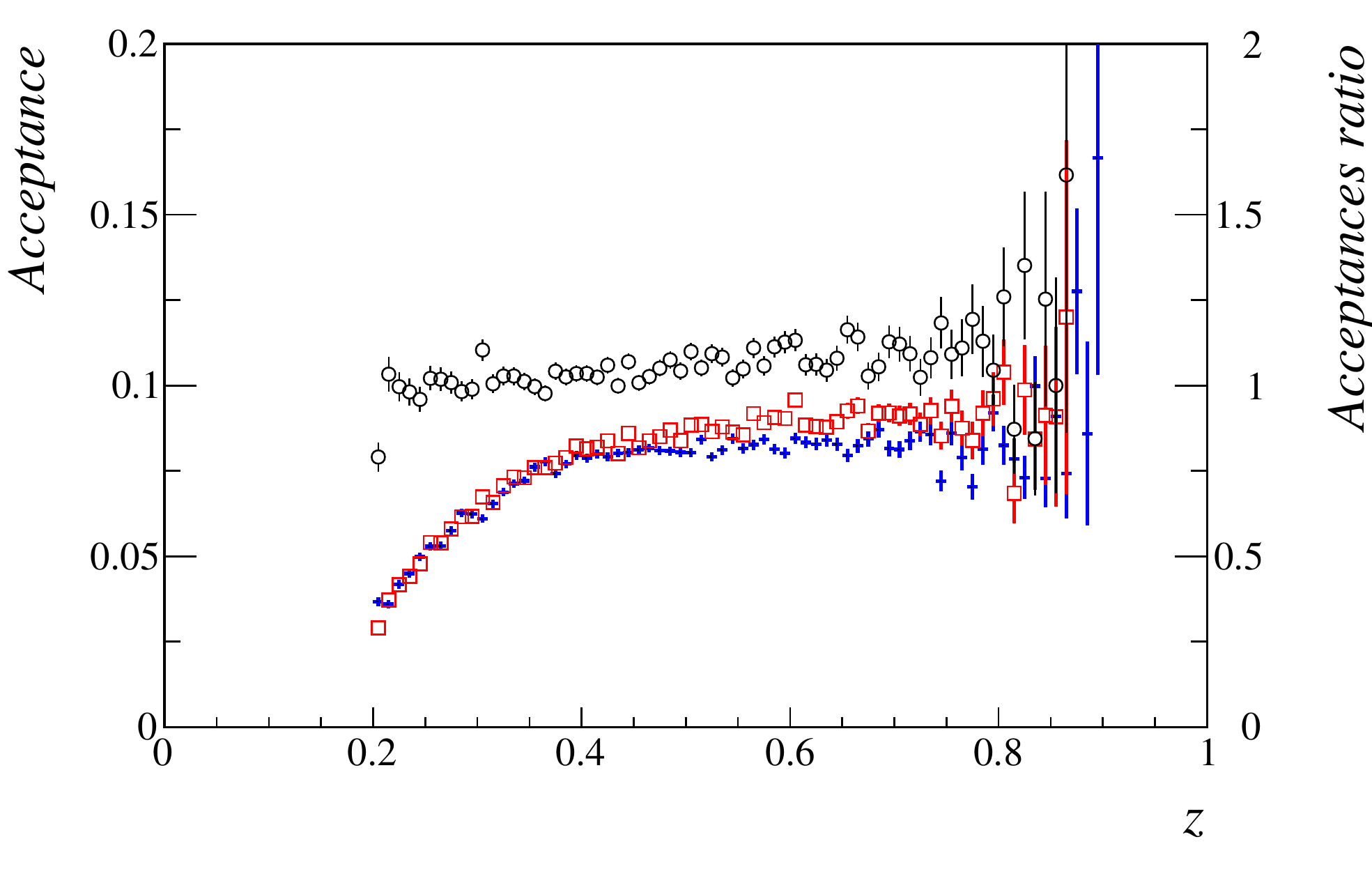}{35}{50}{(c)} \hfill
\OPicTwo{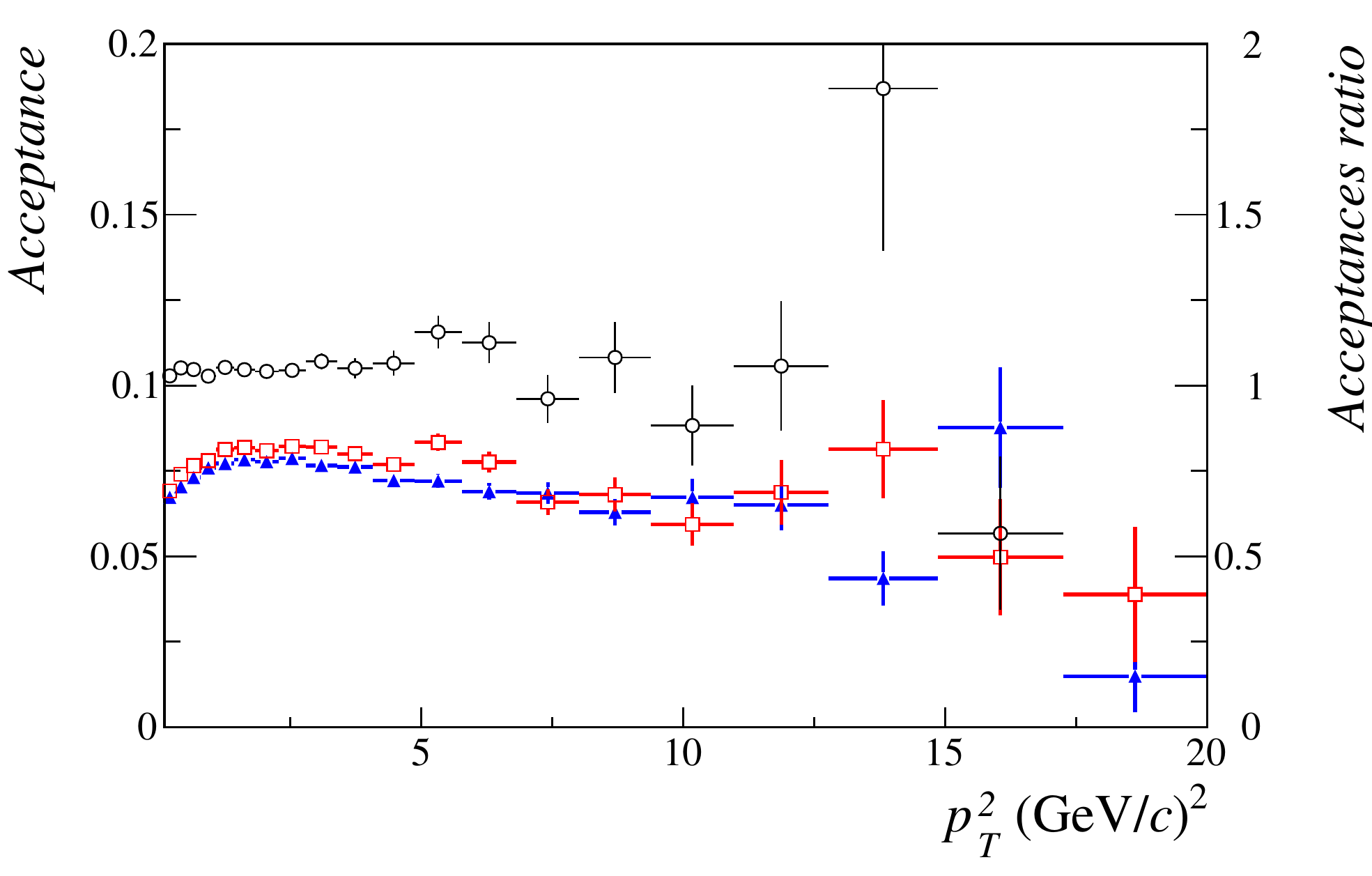}{35}{50}{(d)}
\caption{One-dimensional acceptances for $\DSpm$ production as a function of 
(a) $\nu$, 
(b) $E $, 
(c) $z$ and 
(d) $p_T ^2$.
Red boxes correspond to $\DSp$, 
blue triangles to $\DSm$ events.
The black circles 
show the ratio of acceptances $A_{\DSp} / A_{\DSm}$, 
the ordinates for the ratios are 
drawn on the right-hand side of the figures.
\label{fig:MC:acc:D0 z} }
\end{figure}

The one-dimensional acceptance functions $A_{\DSp}$ and $A_{\DSm}$ are shown in
Fig.~\ref{fig:MC:acc:D0 z} as a function of $\nu$, $E$, $z$ and $p_T ^2$. In
addition, the ratio of the $\DSp$ acceptance over that of $\DSm$ is shown in
each case. Note that within the statistical uncertainty of the Monte Carlo
simulation the acceptances are almost equal for $\DSp$ and $\DSm$, with the
acceptance for $\DSp$ being slightly higher than that for $\DSm$.

The integrated luminosity $\cal L$ is determined by a comparison of the measured number
of inclusive inelastic muon scattering events with the best available measurement of the
corresponding cross-section. The differential number of events is the product
of integrated luminosity, inclusive muon acceptance and inclusive differential 
cross-section:
\begin{equation}
\label{eq:N_L_CS_A}
\frac{d^2 N(x_{Bj},y)}{dx_{Bj} \ dy} = {\cal L} \cdot
A_{\mu}(x_{Bj},y) \cdot \frac{d^2\sigma^{\mu N \rightarrow
\mu'X}(x_{Bj},y)}{dx_{Bj} \ dy}\,.
\end{equation}
The inclusive inelastic muon-deuteron cross-section was measured by the NMC
Collaboration for various muon energies between 90 and 280 GeV and published
as a parameterization of the structure function $F_2$
\cite{F2PARAM}. Thus the cross-section has
to be reconstructed based on this $F_2$ parameterization. The measured
cross-section is connected with the one-photon exchange cross-section via a
radiative correction factor $\eta(x_{Bj},y)$:
\begin{equation}
\label{eq:cs mu-N}
\frac{d^2\sigma^{\mu N \rightarrow \mu'X}(x_{Bj},y)}{dx_{Bj} \ dy}
=\frac{1}{\eta(x_{Bj},y)}\frac{d^2\sigma_{1 
\gamma}(x_{Bj},y)}{dx_{Bj} \ dy}\,.
\end{equation}
The one-photon exchange cross-section is connected with $F_2$ by :
\begin{equation}
\label{eq:cs 1photon}
\frac{d^2\sigma_{1\gamma}(x_{Bj},y)}{dx_{Bj} \ dy} =
\frac{4\pi(\alpha\hbar c)^2}{x_{Bj}yQ^2}\left(1-y-\frac{Q^2}{4E_{\mu}^2}+
\frac{(1-2m^2/Q^2)(y^2+Q^2/E_{\mu}^2)}{2(1+R(x_{Bj},Q^2))}\right)
F_2(x_{Bj},Q^2)\,,
\end{equation}
where $m$ is the muon mass.
The factor $R(x_{Bj},Q^2)$ is the cross-section ratio for longitudinal over
transverse photons:
\begin{equation}
R(x_{Bj},Q^2) = \frac{\sigma_L}{\sigma_T}\,.
\end{equation}
The radiative correction factor $\eta(x_{Bj},y)$ is calculated with codes based
on~\cite{akh96}. The ratios $R(x_{Bj},Q^2)$ are determined as in
Ref.~\cite{Ageev:2005pq}. Given the light material composing the target (Li, D
and He), nuclear effects have been neglected.

The integrated luminosity is determined in bins of $(x_{Bj},y)$ as:
\begin{equation}
\label{eq:L extract}
{\cal L} = \frac{1}{A_{\mu}(x_{Bj},y)}\cdot
\frac{d^2 N(x_{Bj},y)/(dx_{Bj} \ dy)}
{d^2\sigma^{\mu N \rightarrow \mu'X}(x_{Bj},y)/(dx_{Bj} \ dy)} \,.
\end{equation}
The integrated luminosity on the left-hand side of Eq.~\ref{eq:L extract} has to
be constant, while all terms on the right-hand side depend on $x_{Bj}$ and $y$.
As a side product of extracting the integrated luminosity, this equation can be
used to evaluate the uncertainty of the muon acceptance calculation for $Q^2$
values larger than about 0.6~\gomt, where the NMC parameterization is valid.
The values of ${\cal L}$ obtained for different $(x_{Bj},y)$ bins vary indeed by
up to 20\% over the relevant $(x_{Bj},y)$ range, so that an overall systematic
uncertainty of 20\% is attributed to the product of integrated luminosity and
inclusive muon acceptance. The average value of the integrated luminosity is
calculated as a weighted mean of the luminosities determined in $(x_{Bj},y)$
bins, using the data at $Q^2 > 0.6$ \gomt. For a given bin the weight is the
number of events in that bin.
%The average value of the
%integrated luminosity is determined by averaging the luminosities per bin,
%weighted by the number of events in the given bin, over all bins in $x_{Bj}$ and
%$y$ for $Q^2$ values larger than 0.6 \gomt.  
The result for the integrated luminosity of the 2004 data is $0.71 \pm 0.14$
fb$^{-1}$. Since the statistical uncertainty is negligible, only the 
systematic one is quoted.

\section{$\DSpm$ production cross-sections} 
\label{sec:cross section}

The acceptance uncorrected distributions presented in Sec.~\ref{kinedis} were
given for all data taken in 2002-2006.  The signals for $\DZ$ and $\barDZ$ were
summed up, and so were those for $\DSp$ and $\DSm$. In the following, the
semi-inclusive differential cross-sections for $\DSpm$ production, determined
for data from the year 2004 only, will be obtained separately for $\DSp$ and
$\DSm$. The acceptances, the integrated luminosity and the known branching ratio
(2.6\%) of $\DS$ to $K\pi\pi$ are taken into account. At the end of this
section, $\DSp$ and $\DSm$ asymmetries will be shown for all 2002 to 2006 data,
since integrated luminosity and also the acceptances cancel in these asymmetries
to a good approximation.

\begin{figure}
\centering
\OPicTwo{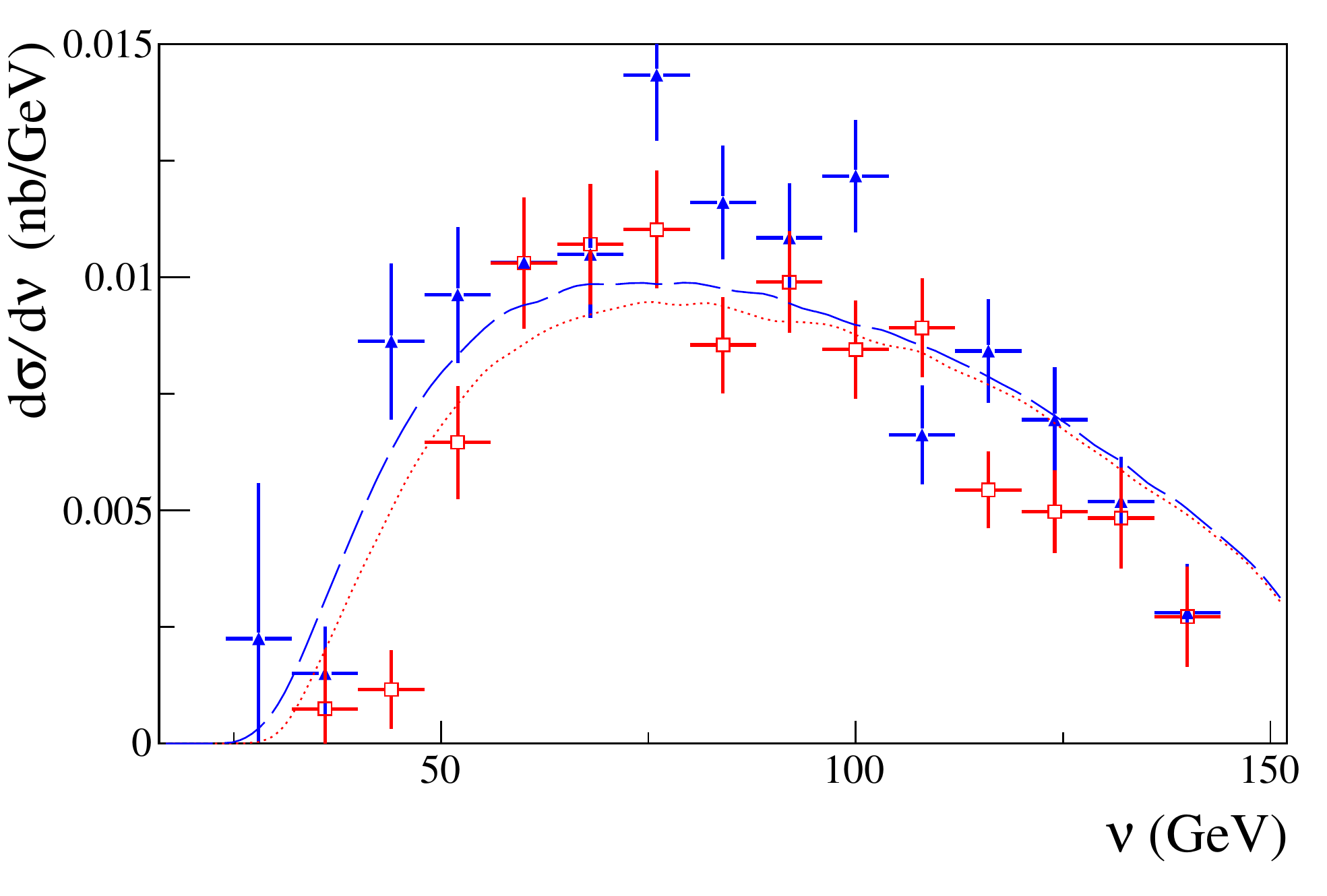}{75}{50}{(a)}  \hfill
\OPicTwo{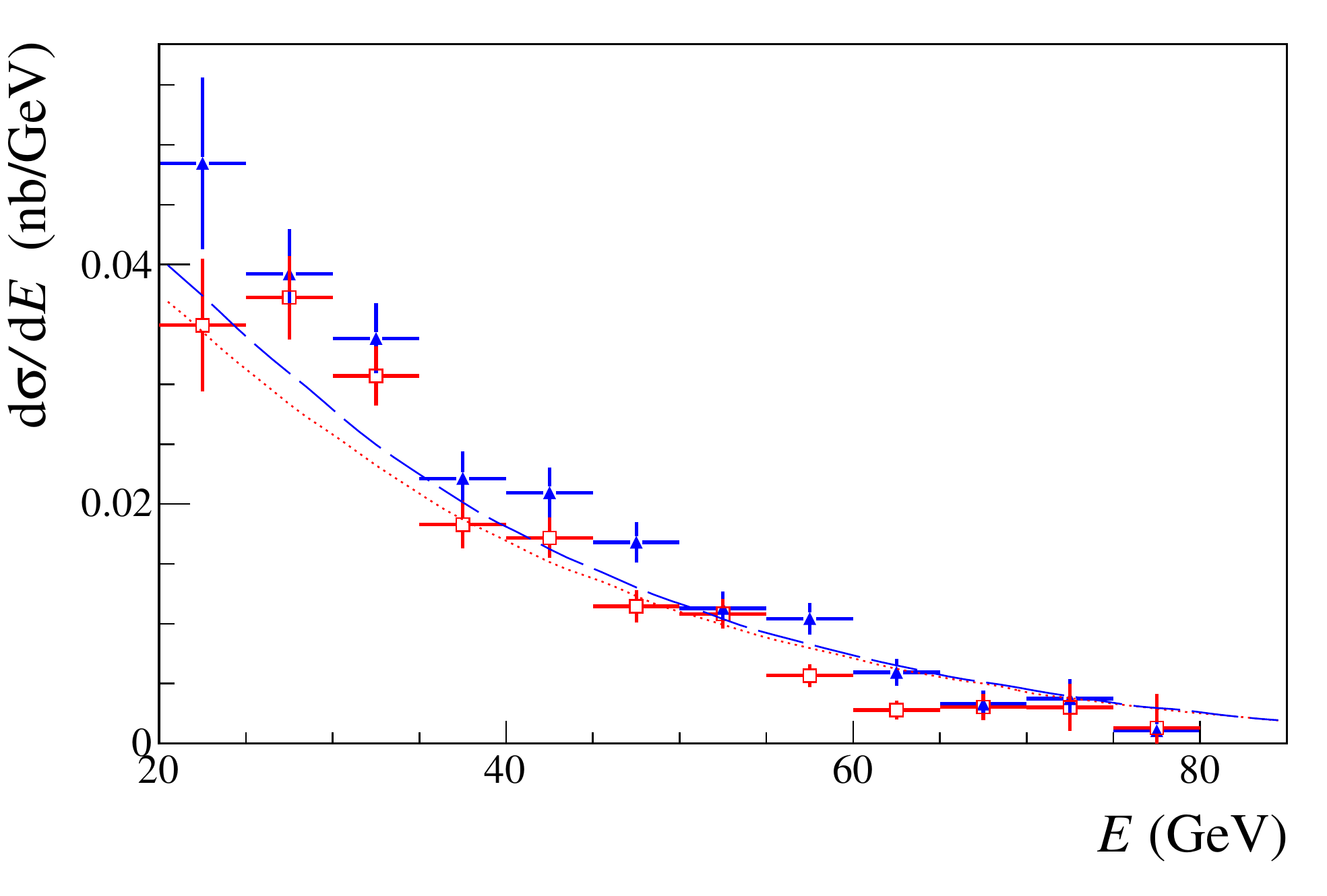}{75}{50}{(b)}  \\
\OPicTwo{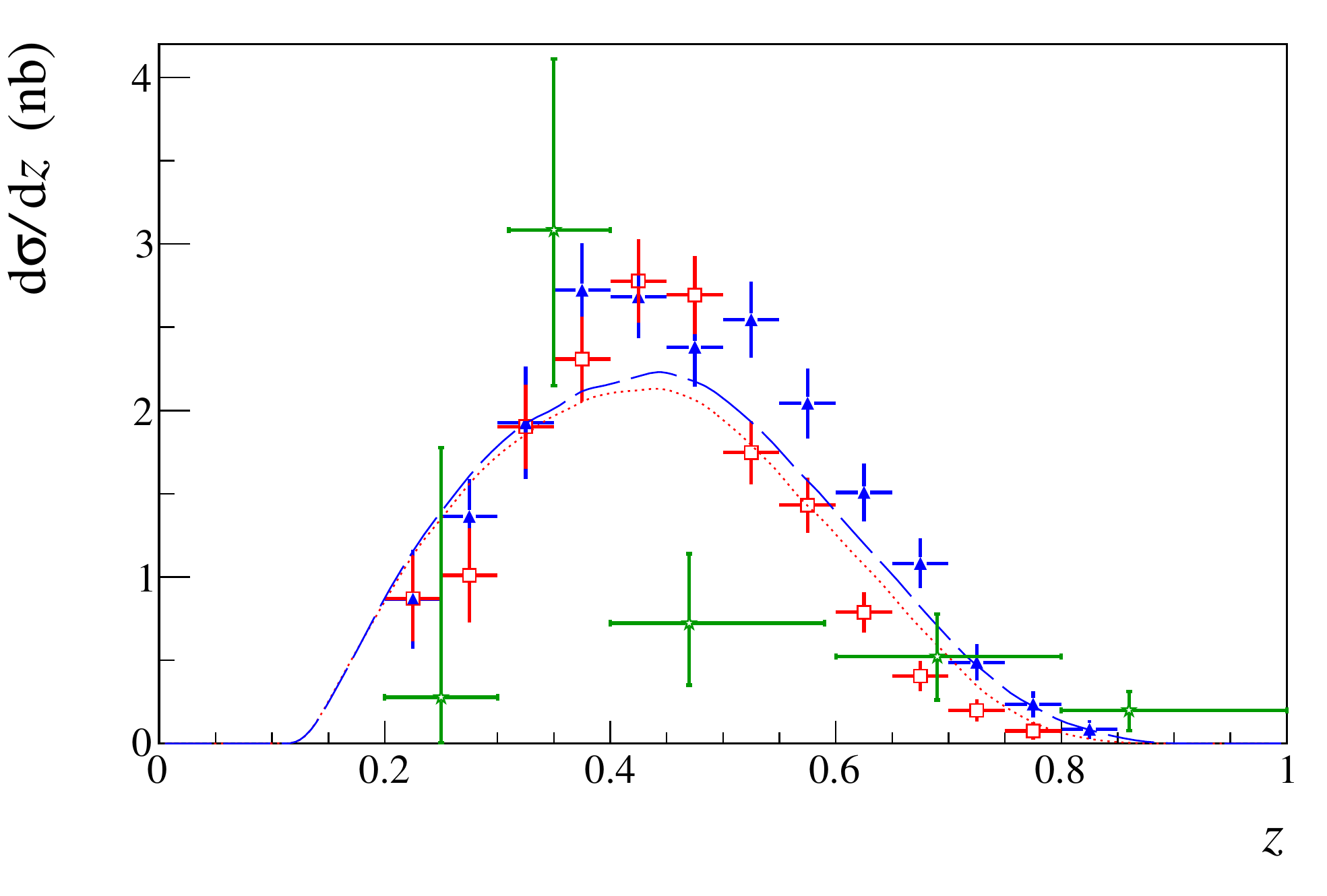}{75}{50}{(c)} \hfill
\OPicTwo{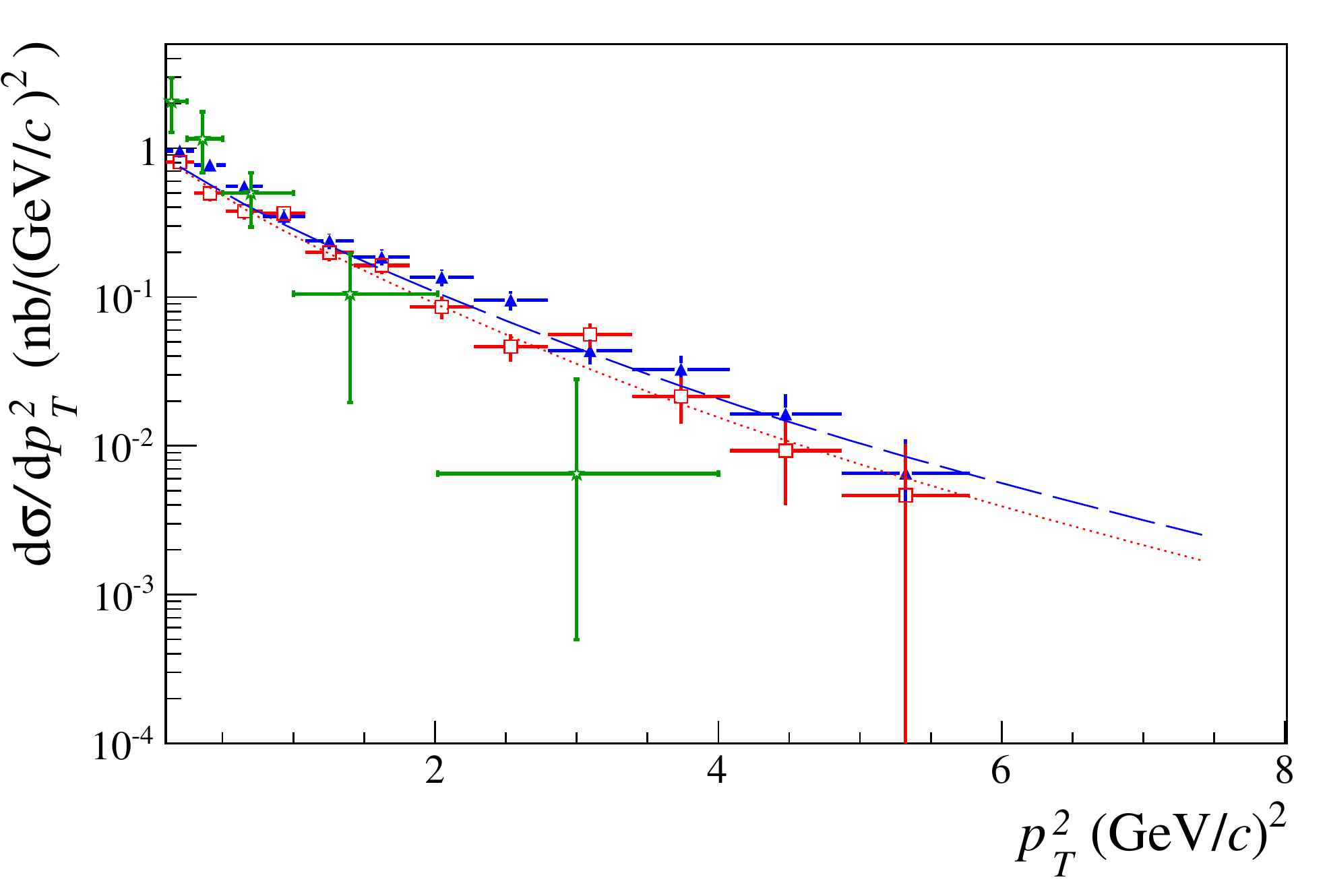}{75}{50}{(d)}  
\caption{Semi-inclusive differential cross-sections for $\DSp$ and $\DSm$
production as a function of (a) virtual photon energy $\nu$, (b) $\DZ$ energy
$E$, (c) fractional energy $z$ and (d) squared transverse momentum $p_T ^2$.
For all distributions, the red squares correspond to $\DSp$ and blue triangles
to $\DSm$ events (2004 data, $\DS$ sample).  The green circles are
semi-inclusive differential cross-sections for $\DZ$ from the EMC
experiment, see text. The curves represent AROMA predictions, dashed for $\DSm$
and dotted for $\DSp$. \label{fig:acc_corr:D0 E}}
\end{figure}

Figure~\ref{fig:acc_corr:D0 E} displays the semi-inclusive differential
cross-sections of $\DSp$ and $\DSm$ events as a function of $\nu, E, z$ and $
p_T^2$. The numerical values of the measured differential cross-sections are
compiled in Table \ref{tab:cs}. These cross-sections are compared with the
theoretical predictions obtained from the AROMA generator, which assumes
$c\bar{c}$ production via photon gluon fusion and includes parton showers.  The
AROMA total cross-section is rescaled to the value of 1.9 nb measured by COMPASS, see
below.  They were calculated using the same program package parameters as those
for the determination of acceptances.  The $ p_T^2$ and $z$ distributions are
also compared with results published by the EMC Collaboration 20 years ago
\cite{EMC}, based on 92 events, obtained with higher muon beam energy and a cut
on $Q^2 >$ 3~\gomt. EMC combined $\barDZ$ and $\DZ$ as within the statistical
precision no differences were observed.  In order to compare with the present
data, their measured values and uncertainties are divided by a factor of 2.

\begin{table}
\caption{
Semi-inclusive differential cross-sections for $\DSp$ and $\DSm$ production as a
function of (a) $\gamma^*$ energy $\nu$, (b) $\DZ$ energy $E$, (c) fractional
energy $z$ and (d) squared transverse momentum $p_T ^2$ of the $\DZ$.  The
central values and bin sizes of $ \nu $ and $ E $ are given in units of GeV,
those of $p_T^2 $ in \gomt.  The last two lines show the integrated
cross-sections. Statistical uncertainties are given. \label{tab:cs}}
\centering
\subfloat[ ]{
\footnotesize
\begin{tabular}{|c|c|c|} \hline
 $ \nu\pm\Delta\nu/2\ \ $ & \multicolumn{2}{c|}{$ \Delta\sigma =(d\sigma/d\nu )\cdot \Delta \nu \ \ \mrf{[nb]} $} \\
 & $\DSp$ & $\DSm$ \\ \hline
$28 \pm 4$ & $ 0.000 \pm 0.000$ & $ 0.018 \pm 0.027$ \\
$36 \pm 4$ & $ 0.006 \pm 0.010$ & $ 0.012 \pm 0.008$ \\
$44 \pm 4$ & $ 0.009 \pm 0.007$ & $ 0.070 \pm 0.014$ \\
$52 \pm 4$ & $ 0.052 \pm 0.010$ & $ 0.078 \pm 0.012$ \\
$60 \pm 4$ & $ 0.083 \pm 0.011$ & $ 0.083 \pm 0.011$ \\
$68 \pm 4$ & $ 0.087 \pm 0.010$ & $ 0.085 \pm 0.011$ \\
$76 \pm 4$ & $ 0.089 \pm 0.010$ & $ 0.116 \pm 0.011$ \\
$84 \pm 4$ & $ 0.069 \pm 0.008$ & $ 0.094 \pm 0.010$ \\
$92 \pm 4$ & $ 0.080 \pm 0.009$ & $ 0.088 \pm 0.010$ \\
$100 \pm 4$ & $ 0.068 \pm 0.009$ & $ 0.098 \pm 0.010$ \\
$108 \pm 4$ & $ 0.072 \pm 0.009$ & $ 0.054 \pm 0.009$ \\
$116 \pm 4$ & $ 0.044 \pm 0.007$ & $ 0.068 \pm 0.009$ \\
$124 \pm 4$ & $ 0.040 \pm 0.007$ & $ 0.056 \pm 0.009$ \\
$132 \pm 4$ & $ 0.039 \pm 0.009$ & $ 0.042 \pm 0.008$ \\
$140 \pm 4$ & $ 0.022 \pm 0.009$ & $ 0.023 \pm 0.009$ \\
\hline
$\sigma^{\DSp}$ , $\sigma^{\DSm}$ & $ 0.762 \pm 0.034$ & $ 0.985 \pm 0.046$ \\
$\sigma^{\DSpm}$ & \multicolumn{2}{c|}{$ 1.747 \pm 0.057$} \\
\hline
\end{tabular}
}
\subfloat[ ] {%
\footnotesize
\centering
\begin{tabular}{|c|c|c|} \hline
 $ E \pm \Delta E/2\ \ $ & \multicolumn{2}{c|}{$ \Delta \sigma= (d\sigma/dE) \cdot \Delta E\ \ \mrf{[nb]} $} \\
 & $\DSp$ & $\DSm$ \\ \hline
$22.5 \pm 2.5$ & $ 0.177 \pm 0.028$ & $ 0.245 \pm 0.036$ \\
$27.5 \pm 2.5$ & $ 0.188 \pm 0.018$ & $ 0.198 \pm 0.019$ \\
$32.5 \pm 2.5$ & $ 0.155 \pm 0.013$ & $ 0.171 \pm 0.015$ \\
$37.5 \pm 2.5$ & $ 0.092 \pm 0.010$ & $ 0.112 \pm 0.012$ \\
$42.5 \pm 2.5$ & $ 0.087 \pm 0.009$ & $ 0.106 \pm 0.011$ \\
$47.5 \pm 2.5$ & $ 0.058 \pm 0.007$ & $ 0.085 \pm 0.009$ \\
$52.5 \pm 2.5$ & $ 0.055 \pm 0.006$ & $ 0.057 \pm 0.007$ \\
$57.5 \pm 2.5$ & $ 0.029 \pm 0.005$ & $ 0.053 \pm 0.007$ \\
$62.5 \pm 2.5$ & $ 0.014 \pm 0.004$ & $ 0.030 \pm 0.006$ \\
$67.5 \pm 2.5$ & $ 0.015 \pm 0.005$ & $ 0.017 \pm 0.006$ \\
$72.5 \pm 2.5$ & $ 0.015 \pm 0.010$ & $ 0.019 \pm 0.008$ \\
$77.5 \pm 2.5$ & $ 0.007 \pm 0.014$ & $ 0.005 \pm 0.007$ \\
& & \\
& & \\
& & \\
\hline
$\sigma^{\DSp}$ , $\sigma^{\DSm}$ & $ 0.892 \pm 0.044$ & $ 1.098 \pm 0.050$ \\
$\sigma^{\DSpm}$ & \multicolumn{2}{c|}{$ 1.990 \pm 0.066$} \\
\hline
\end{tabular}
} \\
\subfloat[ ] {%
\footnotesize
\centering
\begin{tabular}{|c|c|c|} \hline
 $ z \pm \Delta z/2 $ & \multicolumn{2}{c|}{$ \Delta \sigma = (d\sigma/dz) \cdot \Delta z\ \ \mrf{[nb]} $} \\
 & $\DSp$ & $\DSm$ \\ \hline
$0.225 \pm 0.025$ & $ 0.044 \pm 0.013$ & $ 0.044 \pm 0.015$ \\
$0.275 \pm 0.025$ & $ 0.051 \pm 0.014$ & $ 0.069 \pm 0.011$ \\
$0.325 \pm 0.025$ & $ 0.096 \pm 0.013$ & $ 0.097 \pm 0.017$ \\
$0.375 \pm 0.025$ & $ 0.117 \pm 0.013$ & $ 0.138 \pm 0.014$ \\
$0.425 \pm 0.025$ & $ 0.140 \pm 0.013$ & $ 0.136 \pm 0.013$ \\
$0.475 \pm 0.025$ & $ 0.136 \pm 0.012$ & $ 0.120 \pm 0.012$ \\
$0.525 \pm 0.025$ & $ 0.088 \pm 0.010$ & $ 0.129 \pm 0.011$ \\
$0.575 \pm 0.025$ & $ 0.072 \pm 0.008$ & $ 0.103 \pm 0.011$ \\
$0.625 \pm 0.025$ & $ 0.040 \pm 0.006$ & $ 0.076 \pm 0.009$ \\
$0.675 \pm 0.025$ & $ 0.020 \pm 0.005$ & $ 0.055 \pm 0.008$ \\
$0.725 \pm 0.025$ & $ 0.010 \pm 0.003$ & $ 0.025 \pm 0.005$ \\
$0.775 \pm 0.025$ & $ 0.004 \pm 0.003$ & $ 0.012 \pm 0.004$ \\
     &      &      \\

\hline
$\sigma^{\DSp}$ , $\sigma^{\DSm}$ & $ 0.820 \pm 0.035$ & $ 1.008 \pm 0.040$ \\
$\sigma^{\DSpm}$ & \multicolumn{2}{c|}{$ 1.827 \pm 0.053$} \\
\hline
\end{tabular}
}
\subfloat[ ] {%
\hfill
\footnotesize
\centering
\begin{tabular}{|c|c|c|} \hline  
 $ p_{T}^{2} \pm\Delta p_{T}^{2}/2\ $ & \multicolumn{2}{c|}{$ d\sigma/dp_{T}^{2}\ \ \mrf{[nb}/\gomt] $} \\
 & $\DSp$ & $\DSm$ \\ \hline
$ 0.10 \pm 0.10$ & $ 0.865 \pm 0.086$ & $ 1.109 \pm 0.104$ \\
$ 0.31 \pm 0.11$ & $ 0.679 \pm 0.065$ & $ 0.734 \pm 0.075$ \\
$ 0.56 \pm 0.13$ & $ 0.447 \pm 0.052$ & $ 0.610 \pm 0.059$ \\
$ 0.84 \pm 0.15$ & $ 0.350 \pm 0.041$ & $ 0.389 \pm 0.043$ \\
$ 1.16 \pm 0.17$ & $ 0.246 \pm 0.028$ & $ 0.285 \pm 0.032$ \\
$ 1.53 \pm 0.20$ & $ 0.151 \pm 0.022$ & $ 0.200 \pm 0.024$ \\
$ 1.96 \pm 0.23$ & $ 0.115 \pm 0.016$ & $ 0.127 \pm 0.017$ \\
$ 2.45 \pm 0.26$ & $ 0.043 \pm 0.010$ & $ 0.109 \pm 0.014$ \\
$ 3.01 \pm 0.30$ & $ 0.057 \pm 0.010$ & $ 0.059 \pm 0.010$ \\
$ 3.66 \pm 0.35$ & $ 0.023 \pm 0.006$ & $ 0.030 \pm 0.007$ \\
$ 4.40 \pm 0.40$ & $ 0.011 \pm 0.005$ & $ 0.014 \pm 0.005$ \\
$ 5.25 \pm 0.46$ & $ 0.005 \pm 0.011$ & $ 0.012 \pm 0.006$ \\
$ 7.35 \pm 0.60$ & $ 0.002 \pm 0.004$ & $ 0.002 \pm 0.022$ \\
\hline
$\sigma^{\DSp}$ , $\sigma^{\DSm}$ & $ 0.834 \pm 0.036 $ & $ 1.037 \pm 0.048 $ \\
$\sigma^{\DSpm}$ & \multicolumn{2}{c|}{$1.871 \pm 0.060$} \\
\hline
\end{tabular}
}
\end{table}

Good agreement is observed between the shapes of the measured distributions and
the corresponding AROMA predictions.  The distributions of $\DSp$ and $\DSm$ as
a function of $\nu$ show that the points for $\DSm$ are systematically higher
than those for $\DSp$. The effective threshold of $\DSp$ appears to be about
10~\gmass\ higher than that of $\DSm$. The AROMA generator produces also
somewhat more $\DSm$ than $\DSp$ but the differences at threshold are far less
pronounced.  A similar feature can be observed for the $z$ distribution.
In the large-$z$ region, that has a large contribution from low-$\nu$ events,
the cross-section of $\DSm$ becomes significantly larger than that of $\DSp$.
The AROMA calculations predicts more $\DSm$ than $\DSp$ as well,  but the size of the
effect is smaller. For the semi-inclusive differential cross-sections as a
function of $E$ and $p_T^2$, no remarkable differences are observed between
the shapes of the distributions of $\DSp$ and $\DSm$.

The total cross-sections for $\DSp$, $\DSm$ and $\DSpm$ production are extracted
by integration of the differential ones. The differences between the
results from the integration over $\nu$, $E$, $z$ and $p_T^2$ (see Table
\ref{tab:cs}) are used to evaluate the systematic uncertainty of acceptance
corrections.  Using the RMS of the four results (from $\nu$, $E$, $z$ and
$p_T^2$) one obtains a systematic contribution of 0.05 for both $\DSp$ and
$\DSm$ and 0.10 for the sum $\DSpm$, i.e.\ at the level of the statistical
uncertainty.  In the ratio of $\DSp$ over $\DSm$  the acceptances  almost
cancels. The values of the ratio vary between 0.77 and 
0.81, with an average of 0.80 and a RMS of $<0.02$, i.e.\ two to three times
smaller than the statistical uncertainty of $\sim 0.05$.

The final result for the $\DS$ meson production cross-section is then $\sigma(
\mu N \ra \mu ^{\prime} \DSpm X) = 1.86 \pm 0.06\ (\mrf{stat}) \pm 0.10\
(\mrf{sys}) \pm 0.37\ (\mrf{luminosity})$ nb.  The only cut applied is the
energy window for the $\DZ$ meson between $20\ \mrf{GeV} < E < 80\ \mrf{GeV}$ in
the laboratory frame, corresponding to $22\ \mrf{GeV} < E < 86\ \mrf{GeV}$ for
the $\DS$ energy.

For charm-anticharm production, AROMA gives a cross-section of 7.2 nb with
1.35~\gmass\ chosen as the default charm quark mass.  Using the common assumption 
of 0.6 $\DS$ mesons per charm event and accounting for the energy cut $20\
\mrf{GeV} < E < 80\ \mrf{GeV}$, which reduces the number of charm Monte Carlo
events by another factor of 0.6, the corresponding AROMA cross-section predicted
for COMPASS is 2.6 nb. Given the number of assumptions which underlie the AROMA
default options (charm quark mass, fragmentation, no radiative corrections,
leading order QCD apart from parton showers) the agreement with the above
experimental result is considered to be good.

However, deviations from the AROMA predictions are observed in the
data with respect to $\DSp$ and $\DSm$ production. These may provide valuable
insight into their production mechanisms. In a simple LO approach, assuming
photon-gluon fusion with independent fragmentation of the charm and anti-charm
quarks to be the relevant production mechanism, no differences should be
observed between $\DSp$ and $\DSm$. Differences may occur for all
processes where the quark content of the target nucleon matters. 
The quark content of $\DS$ mesons indicate that only $\DSm$ may 
contain a valence quark from the target nucleon.
Furthermore, instead of fragmenting into $\DSp$
the $c$ quark, together with a diquark of the target nucleon, can hadronize into
a charmed baryon, leading to associated production of e.g.\ $\DSm
\Lambda_c$. Thus the $\DSm$ may result from a valence quark and/or associated
production. If parton showers are included in AROMA the flavour dependent quark
distribution functions of the nucleon come into play. Processes like associated
production of $\DSm \Lambda_c$ lead to differences between kinematic
distributions of $\DSp$ and $\DSm$. The same happens for processes where an
initial quark in the nucleon absorbs the virtual photon and radiates a heavy
gluon which then decays to $\ccbar$, or where in the course of fragmentation the
$\cbar$ quark picks up quarks from the nucleon.

\begin{figure}
\centering
\OPicTwo{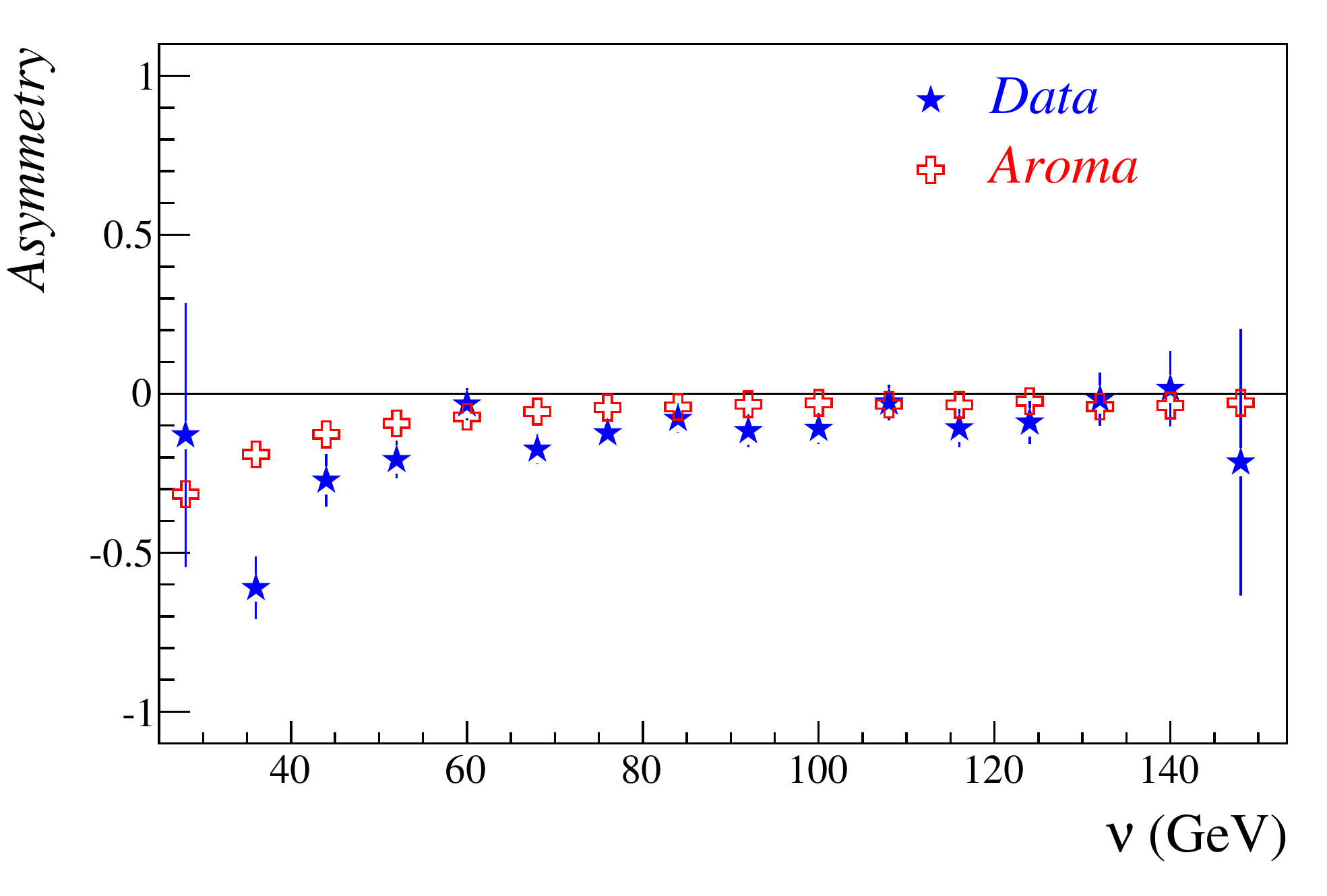}{25}{50}{(a)}  \hfill
\OPicTwo{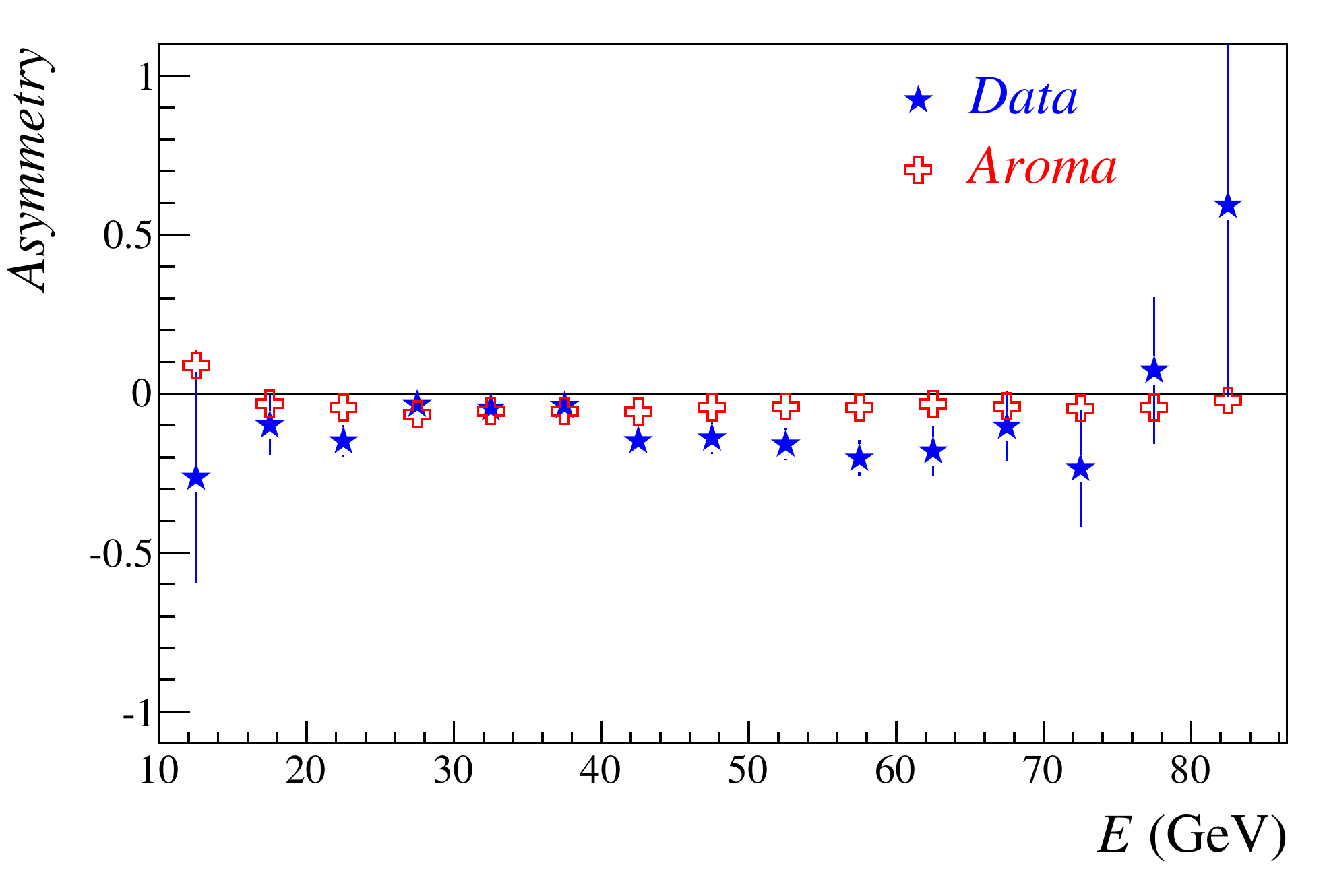}{25}{50}{(b)}  \\
\OPicTwo{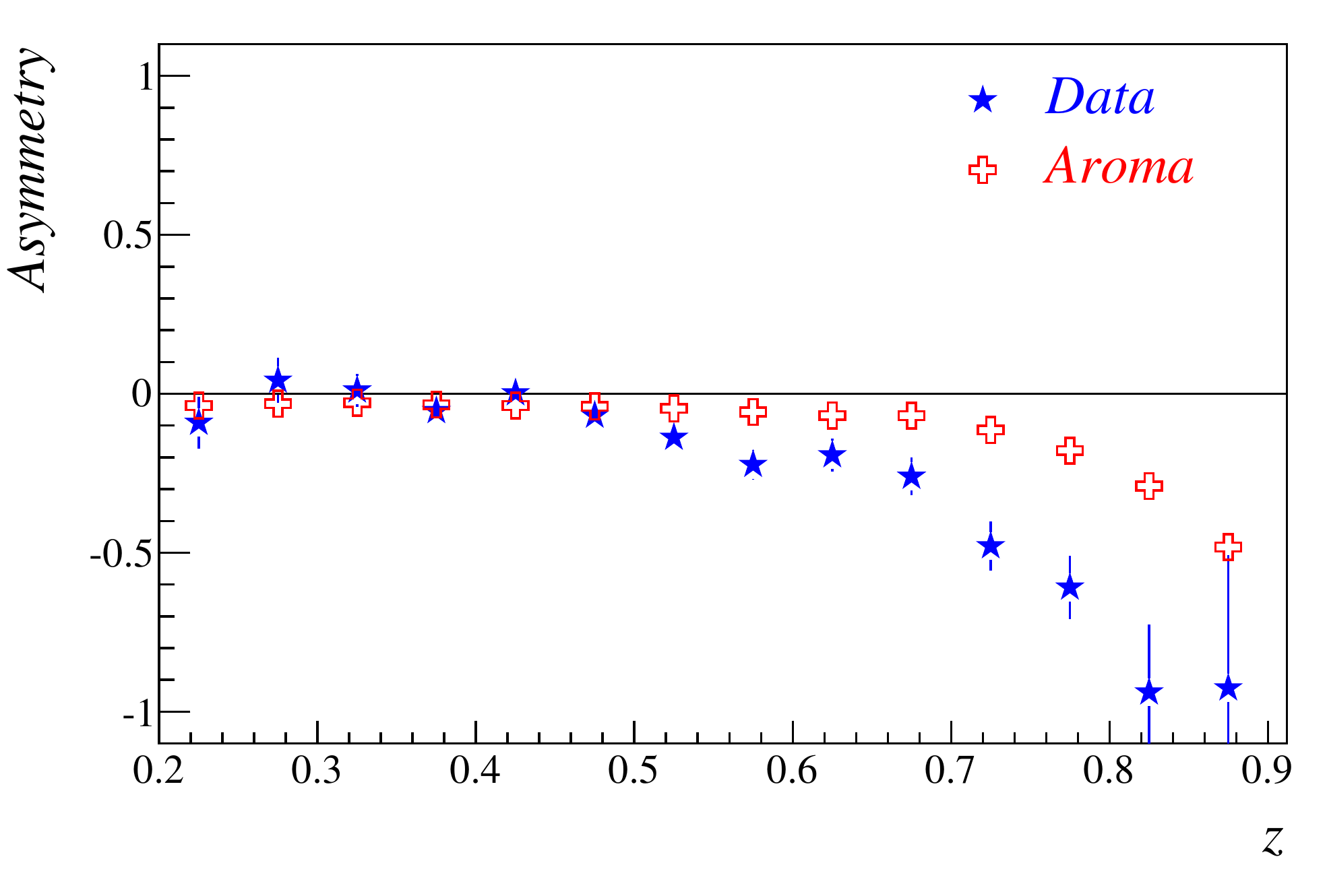}{25}{50}{(c)}  \hfill
\OPicTwo{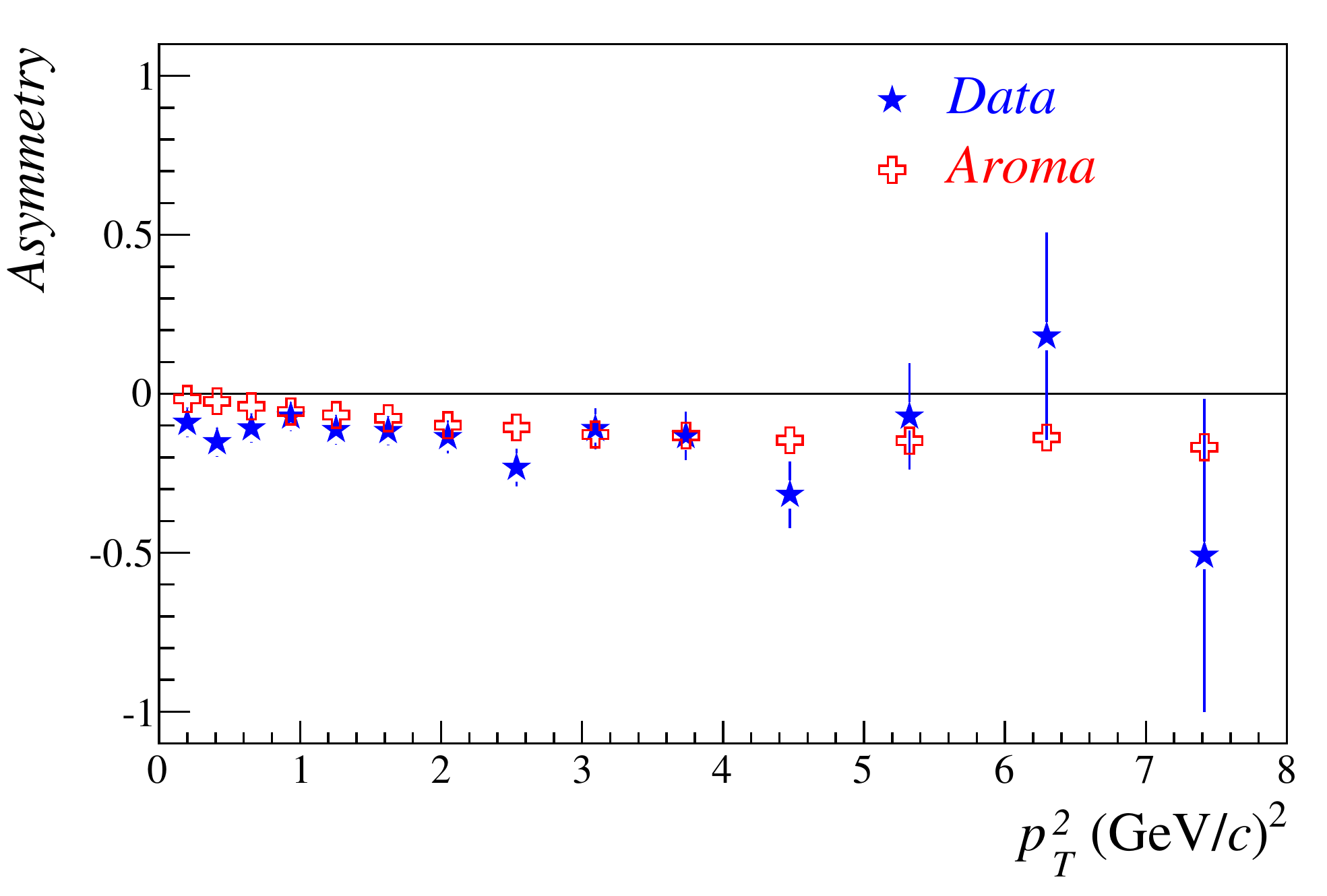}{25}{50}{(d)} 
\caption{ Measured $\DSp$ and $\DSm$ asymmetries for data (blue stars) and AROMA
generator (red crosses) events as a function of $X = \nu,\ E,\ z$ and $p_T^2 $
(coloured online).  All 2002-2006 data are used. \label{fig:zD0_diff}
 }
\end{figure}

\begin{table}
\caption{ Measured asymmetry $ A(X)$
as a function of $X = \nu,\ E,\ z$ and $p_T^2 $. The central values and bin sizes 
of $\nu$ and $ E$ in (a) and (b) are given in units of GeV, 
those of $p_T^2 $ in (d) in units of \gomt. \label{tab:asym nu} }
 \centering
 \subfloat[ ]{
  \footnotesize
  \begin{tabular}{|c|c|c|}
  \hline
   $ \nu\pm\Delta\nu/2\ \ $ & $ A(\nu )$\\
  \hline
   $ 28 \pm 4$ & $-0.130 \pm 0.415$ \\
   $ 36 \pm 4$ & $-0.610 \pm 0.098$ \\
   $ 44 \pm 4$ & $-0.272 \pm 0.082$ \\
   $ 52 \pm 4$ & $-0.207 \pm 0.059$ \\
   $ 60 \pm 4$ & $-0.032 \pm 0.051$ \\
   $ 68 \pm 4$ & $-0.174 \pm 0.047$ \\
   $ 76 \pm 4$ & $-0.123 \pm 0.044$ \\
   $ 84 \pm 4$ & $-0.078 \pm 0.046$ \\
   $ 92 \pm 4$ & $-0.116 \pm 0.053$ \\
   $100 \pm 4$ & $-0.109 \pm 0.048$ \\
   $108 \pm 4$ & $-0.027 \pm 0.057$ \\
   $116 \pm 4$ & $-0.108 \pm 0.060$ \\
   $124 \pm 4$ & $-0.090 \pm 0.068$ \\
   $132 \pm 4$ & $-0.018 \pm 0.084$ \\
   $140 \pm 4$ & $+0.016 \pm 0.119$ \\
   $148 \pm 4$ & $-0.215 \pm 0.420$ \\
  \hline
  \end{tabular}
 }
 \subfloat[ ]{%
  \footnotesize
  \begin{tabular}{|c|c|c|}
  \hline
   $ E\pm\Delta E/2\ \ $ & $ A(E)$ \\
  \hline
   $12.5 \pm 2.5$ & $-0.263 \pm 0.333$ \\
   $17.5 \pm 2.5$ & $-0.099 \pm 0.092$ \\
   $22.5 \pm 2.5$ & $-0.149 \pm 0.051$ \\
   $27.5 \pm 2.5$ & $-0.034 \pm 0.041$ \\
   $32.5 \pm 2.5$ & $-0.045 \pm 0.039$ \\
   $37.5 \pm 2.5$ & $-0.038 \pm 0.042$ \\
   $42.5 \pm 2.5$ & $-0.148 \pm 0.043$ \\
   $47.5 \pm 2.5$ & $-0.139 \pm 0.049$ \\
   $52.5 \pm 2.5$ & $-0.159 \pm 0.050$ \\
   $57.5 \pm 2.5$ & $-0.203 \pm 0.057$ \\
   $62.5 \pm 2.5$ & $-0.180 \pm 0.080$ \\
   $67.5 \pm 2.5$ & $-0.103 \pm 0.109$ \\
   $72.5 \pm 2.5$ & $-0.235 \pm 0.185$ \\
   $77.5 \pm 2.5$ & $+0.074 \pm 0.231$ \\
   $82.5 \pm 2.5$ & $+0.593 \pm 0.604$ \\
   $87.5 \pm 2.5$ & $+0.288 \pm 2.146$ \\
  \hline
  \end{tabular}
 } \\
 \subfloat[]{%
  \label{tab:asym z}
  \footnotesize
  \begin{tabular}{|c|c|c|}
  \hline
   $ z\pm \Delta z/2 $ & $ A(z)$\\
  \hline
   $0.225 \pm 0.025$ & $-0.090 \pm 0.082$ \\
   $0.275 \pm 0.025$ & $+0.042 \pm 0.071$ \\
   $0.325 \pm 0.025$ & $+0.011 \pm 0.052$ \\
   $0.375 \pm 0.025$ & $-0.054 \pm 0.043$ \\
   $0.425 \pm 0.025$ & $+0.002 \pm 0.041$ \\
   $0.475 \pm 0.025$ & $-0.068 \pm 0.040$ \\
   $0.525 \pm 0.025$ & $-0.137 \pm 0.041$ \\
   $0.575 \pm 0.025$ & $-0.223 \pm 0.046$ \\
   $0.625 \pm 0.025$ & $-0.193 \pm 0.052$ \\
   $0.675 \pm 0.025$ & $-0.260 \pm 0.059$ \\
   $0.725 \pm 0.025$ & $-0.479 \pm 0.078$ \\
   $0.775 \pm 0.025$ & $-0.609 \pm 0.099$ \\
   $0.825 \pm 0.025$ & $-0.939 \pm 0.213$ \\
   $0.875 \pm 0.025$ & $-0.926 \pm 0.419$ \\
  \hline
  \end{tabular}
 }
 \subfloat[]{%
  \label{tab:asym Pt2}
  \footnotesize
  \begin{tabular}{|c|c|c|}
  \hline
   $ p_{T}^{2}\pm\Delta p_{T}^{2}/2\ \ $ & $A(p_{T}^{2})$ \\
  \hline
  $0.10 \pm 0.10$ & $-0.094 \pm 0.046$ \\
  $0.31 \pm 0.11$ & $-0.105 \pm 0.046$ \\
  $0.56 \pm 0.13$ & $-0.151 \pm 0.045$ \\
  $0.84 \pm 0.15$ & $-0.082 \pm 0.045$ \\
  $1.16 \pm 0.17$ & $-0.083 \pm 0.045$ \\
  $1.53 \pm 0.20$ & $-0.076 \pm 0.044$ \\
  $1.96 \pm 0.23$ & $-0.148 \pm 0.049$ \\
  $2.45 \pm 0.26$ & $-0.192 \pm 0.058$ \\
  $3.01 \pm 0.30$ & $-0.165 \pm 0.061$ \\
  $3.66 \pm 0.35$ & $-0.124 \pm 0.077$ \\
  $4.40 \pm 0.40$ & $-0.208 \pm 0.105$ \\
  $5.25 \pm 0.46$ & $-0.253 \pm 0.155$ \\
  $6.23 \pm 0.52$ & $+0.380 \pm 0.266$ \\
  $7.35 \pm 0.60$ & $-0.520 \pm 0.496$ \\
  \hline
  \end{tabular}
 }
\end{table}

In order to provide statistically more precise information on the potentially
interesting differences between $\DSp$ and $\DSm$ production,
Fig.~\ref{fig:zD0_diff} shows particle-antiparticle asymmetries of the
semi-inclusive cross-sections,
\begin{equation}
A(X) =\frac{d\sigma ^{\DSp}(X) -d\sigma ^{\DSm}(X)}{d\sigma ^{\DSp}(X) + d
\sigma^{\DSm}(X)}
\end{equation}
as a function of $X = \nu,\ E,\ z$ and $p_T^2 $ for both the $\DS$ sample and
Monte Carlo events generated by AROMA. Here the full statistics of the
years 2002-2006 is used. It is assumed that the acceptances for the two charge
combinations are equal. In the previous section it was shown that for the year
2004 this is indeed approximately true.  The numerical values of the measured
asymmetries are given in Table~\ref{tab:asym nu}, where only statistical
uncertainties are shown, based on the assumption that acceptance cancels. A
small cross-section assymmetry between $\DP$ and $\DM$ production has been observed
recenlty in a different energy range  by the LHCb experiment~\cite{LHCb:2012fb}.

As one can see from the figure, the measured asymmetry decreases significantly
stronger than that predicted by AROMA when $\nu$ decreases below 40~GeV and/or
when $z$ increases above 0.6. The distributions shown as a function of $\nu$
clearly exhibit different thresholds for $\DSp$ and $\DSm$ production, which
supports a stronger presence of mechanisms other than PGF with independent
fragmentation. As a function of $z$, the most pronounced differences between
$\DSp$ and $\DSm$ are seen at large values of $z$, whereas at $z$ values lower
than 0.5 the production rates are nearly equal.  Values of $z$ larger than 0.5
indicate an asymmetric sharing of the energies between a $D$ meson and its
associated partner with opposite charm content. Since the cross-section of
$\DSm$, which contains a down and an anti-charm quark, increases with increasing
$z$ stronger than that of $\DSp$, this observation suggests processes where the
anti-charm quark is fast and the charm-quark is slow. Here, a candidate process
is again associated production of a $\DSm$ along with a charmed baryon, i.e
$\DSm$ $\Lambda_c$. Alternatively, since the $\DSm$ may also contain a valence
quark of the nucleon whereas the $\DSp$ does not, one may think of processes
other than associated production, which involve valence quarks of the nucleon.

Asymmetries between the production of $\DZ$ and $\barDZ$ or
$\DSp$ and $\DSm$ were already observed in numerous earlier experiments (see
e.g.~\cite{E687} for charm photoproduction and~\cite{SELEX, SELEX2, WA82, E769,
E791, WA89, HALLING} for charm production by hadrons), although not as
pronounced as in the present experiment that covers the region of virtual-photon
energies from threshold up to 140 GeV.

\section{Summary and conclusions } 

The observed total cross section of (\TotalCrossSection ) nb for the production
of $\DSp$ and $\DSm$ mesons in inelastic muon nucleon interactions at 160 GeV
incident muon energy within the COMPASS acceptance ($20\ \mrf{GeV} < E < 80\
\mrf{GeV}$ and $22\ \mrf{GeV} < E < 86\ \mrf{GeV}$, for $\DZ$ and $\DS$
respectively) lies within the range of values expected if the dominant process
is photon-gluon fusion to open charm production. The total error is dominated by
the uncertainty on the luminosity.  

The detailed comparison of the measured differential cross sections of
$\DS$ production as a function of the variables $\nu$, $E$, $z$ and 
$p_T^2$ shows good agreement with those expected from the model underlying the
AROMA generator used to produce the theoretical distributions. This is
remarkable as most of the kinematic distributions
of $D$ mesons are quite different in shape compared to those of the
background and the neighbouring $\KTS$ resonance.

The observed large asymmetries between $\DSp$ and $\DSm$ production for $\nu <
40$ GeV and $z>0.6$ can only partially be described by the model used in AROMA,
which predicts differences of the same sign but of smaller magnitude. This
indicates that the hadronization processes of charm and anti-charm quarks differ
more significantly than expected or/and processes other than PGF contribute by a
larger amount than assumed.
   
The observed dependences of these differences on the kinematic variables, in
particular on the photon energy $\nu$ and the fractional energy $z$, suggest
that associated production (e.g.\ $D \Lambda_c$) plays a dominant role at low
photon energy. Also, $\DSm$ production involving valence quarks of the nucleon
may contribute to the observed asymmetries between $\DSp$ and $\DSm$
production.

\section*{Acknowledgements}

We gratefully acknowledge the support of the CERN management and staff
and the skill and effort of the technicians of our collaborating
institutes. Special thanks go to V.~Anosov and V.~Pesaro for their
technical support during the installation and the running of this
experiment. This work was made possible by the financial support of
our funding agencies.

\bibliographystyle{utcaps2}
\bibliography{references}

\end{document}